\documentclass[twocolumn]{aastex631}
\usepackage{svg}

\usepackage{CJKutf8}
\usepackage{color}

\usepackage{mathrsfs}
\usepackage{rsfso}
\usepackage{rotating}

\usepackage{graphicx}	% Including figure files
\usepackage{amsmath}	% Advanced maths commands
\usepackage{amssymb}	% Extra maths symbols

\newcommand\lsim{\mathrel{\rlap{\lower4pt\hbox{\hskip1pt$\sim$}}
\raise1pt\hbox{$<$}}}
\newcommand\gsim{\mathrel{\rlap{\lower4pt\hbox{\hskip1pt$\sim$}}
\raise1pt\hbox{$>$}}}

\shortauthors{Faridani et al.}

\graphicspath{{./}{figures/}}

\defcitealias{Morton+16}{Mo16}
\defcitealias{Valizadegan+22}{V22}
\defcitealias{Mayo+18}{Ma18}

\shorttitle{More Likely Than You Think}
\shortauthors{Faridani et al.}

\graphicspath{{./}{figures/}}
\begin{document}

\title[More Likely Than You Think]{More Likely Than You Think: Inclination-Driving Secular Resonances are Common in Known Exoplanet Systems}

\author[0000-0003-3799-3635]{Thea H. Faridani}
\correspondingauthor{Thea H. Faridani}
\email{thfaridani@astro.ucla.edu}
\affiliation{Department of Physics and Astronomy, University of California, Los Angeles, CA 90095, USA}
\affiliation{Mani L. Bhaumik Institute for Theoretical Physics, Department of Physics and Astronomy, UCLA, Los Angeles, CA 90095, USA}

\author[0000-0002-9802-9279]{Smadar Naoz}
\affiliation{Department of Physics and Astronomy, University of California, Los Angeles, CA 90095, USA}
\affiliation{Mani L. Bhaumik Institute for Theoretical Physics, Department of Physics and Astronomy, UCLA, Los Angeles, CA 90095, USA}

\author[0000-0001-8308-0808]{Gongjie Li}
\affiliation{School of Physics, Georgia Institute of Technology, Atlanta, GA 30332, USA}

\author[0000-0002-7670-670X]{Malena Rice}
\affiliation{Department of Astronomy, Yale University, New Haven, CT 06520, USA}

\author[0009-0002-3550-2310]{Nicholas Inzunza}
\affiliation{Department of Physics and Astronomy, University of California, Los Angeles, CA 90095, USA}

\begin{abstract}

Multi-planet systems face significant challenges to detection. For example, further-orbiting planets have reduced signal-to-noise ratio in radial velocity detection methods, and small mutual inclinations between planets can prevent them from all transiting. One mechanism to excite mutual inclination between planets is secular resonance, where the nodal precession frequencies of the planets align such as to greatly increase the efficiency of angular momentum transport between planets. These resonances can significantly misalign planets from one another, hindering detection, and typically can only occur when there are three or more planets in the system. Naively, systems can only be in resonance for particular combinations of planet semimajor axes and masses; however, effects that alter the nodal precession frequencies of the planets, such as the decay of stellar oblateness, can significantly expand the region of parameter space where resonances occur. In this work, we explore known three-planet systems, determine whether they are in (or {were in}) secular resonance {due to evolving stellar oblateness}, and demonstrate the implications of resonance on their detectability and stability. {{We show that about $20\%$ of a sample of three planet transiting systems seem to undergo these resonances early in their lives. }}
\end{abstract}

\section{Introduction}

The multi-planet Kepler systems, often called ``Kepler Multis,'' are a class of exoplanet systems uncovered by the \textit{Kepler} mission \citep[][]{Lissauer+11}. They are characterized by hosting {3 or more} planets with even spacing \citep[e.g.,][]{Weiss+18Peas}, low mutual inclinations \citep[e.g.,][]{Lissauer+11,Fang+12,Fabrycky+14}, low eccentricities \citep[e.g.,][]{VanEylen+15,VanEylen+19}, and a general absence of mean-motion resonant configurations \citep[e.g.,][]{Fabrycky+14}. Moreover, there are suggested correlations in mass and radius of the planets within these systems \citep[{i.e.}, the ``Peas in the pod" pattern;][]{Weiss+18Peas,Millholland+17}. While detection bias may be a contributing factor in the detection of this pattern \citep[e.g.,][]{Murchikova+20,Zhu20}, these biases alone cannot realistically account for all the correlations observed \citep[e.g.,][]{Weiss_2020,Weiss+23}. As these are high-multiplicity exoplanet systems, they experience planet-planet interactions. These interactions have been studied extensively in the literature, {and can sculpt the present-day configurations of exoplanet systems both through short-timescale effects like scattering and long-timescale effects like secular evolution} \citep[see e.g.,][]{LithwickWu+11,VanLaerhoven+12,Naoz+13,Lanza+14,Petrovich+19,Denham+19,Volk+20,wei+21,Lu24}.

As an example of the significance of planet-planet interactions, secular resonances can significantly impact exoplanet systems. Secular resonances can lead to increased rates of angular momentum exchange when certain precessional frequencies align, discussed in detail in section \ref{sec:math}. These frequencies are much slower than the orbital frequency, hence, ``secular". Secular resonances can interact with tides to circularize or migrate planetary orbits \citep[e.g.,][]{Hansen+15}. In systems with {three or more} planets, resonances can overlap, causing chaotic evolution of planets' orbits, known as ``secular chaos,'' which can destabilize systems, produce ultra short-period planets, and pollute white dwarfs \citep[e.g.,][]{LithwickWu+11,Petrovich+19,O'Connor+22}.

The secular planet-planet interactions can be quantified to linear order in eccentricity and inclination by the well-studied ``Laplace-Lagrange'' formalism \citep[see, e.g., Ch. 7 of][]{Murray+00book}. The Laplace-Lagrange formalism models planets and other bodies orbiting a central body (usually a star) as massive rings, which torque and deform each other, changing their eccentricities and inclinations but conserving energy (semimajor axes). Laplace-Lagrange correctly predicts the initial conditions that produce first-order secular resonances (which depend only on the bodies' masses and semimajor axes). Laplace-Lagrange itself cannot predict higher-order secular resonances, such as those that create ``secular chaos'' \citep[see, e.g.,][]{LithwickWu+11} that may only appear for particularly excited eccentricity or inclination values. Beyond planet-planet interaction, in recent years, there has been a growing interest in the star-planet interaction, including interactions arising from general relativity, the evaporating protoplanetary disk, and the stellar gravitational quadrupole moment, $J_2$ \citep[e.g.,][]{Plavchan+15,Faridani+22,Faridani+23,Spalding+16,Becker+20,Brefka+21,wei+21, chen+22}. 

The evolving disk is of particular note because of its time-evolution. Its evolution creates a precession in the planets' longitudes of ascending node and arguments of periapsis, which can bring them into and out of secular resonance. The time-variation of the precession means that the resonance moves across a (potentially quite large) region of the parameter space rather than staying in place. This is commonly called a ``sweeping secular resonance". {If a system is in a region of parameter space where the sweeping resonance occurs, the resonance can} imprint a signature on the orbital elements of the planets post-evaporation. Sweeping resonances have been identified in the evolving disk and have been shown to affect planet formation \citep[e.g.,][]{Best+24} and excite significant eccentricity and inclination in nearby planets post-formation \citep[e.g.,][]{Heppenheimer+80, Ward+81,Nagasawa+03, LiLai+23, Zanazzi+24, Petrovich+20}.

The stellar gravitational quadrupole moment, $J_2$ is a correction to the stellar gravitational potential arising from a star's deviations from spherical symmetry \citep[e.g.,][]{Murray+00book}.
This asymmetry causes nodal recession and periapsis precession for any planets orbiting that star at a rate of
\begin{equation}
    \dot{\Omega}_{J_2},\,\, \dot{\varpi}_{J_2} = \mp n\left[\frac{3}{2} J_{2}\left(\frac{R_{\mathrm{\star}}}{a}\right)^{2}\right] \propto \mp J_2 P^{-7/3},
\end{equation}
where $P$ is an individual planet's orbital period \citep[e.g.,][]{Murray+00book}. 
The quadrupole moment $J_2$ was shown to misalign tightly packed planets from one another \citep[e.g.,][]{Spalding+16,Becker+20,Brefka+21}, impact transit probabilities in single-planet systems \citep[e.g.,][]{Stephan+23} contribute to tidal migration \citep[e.g.,][]{Pu+19,Millholland+20}, and has been used to analyze the orbital inclination of Mercury \citep[e.g.,][]{Ward+76}.

For sun-like, main-sequence stars, the evolution of their spins results in a significant evolution of their $J_2$ values. During their pre-main-sequence phases, sun-like stars have an observed range of spins, from 10-day periods to as fast as one day \citep[e.g.,][]{Gallet+13}. However, it is well-known that main sequence stars lose angular momentum via magnetic braking on their stellar wind \citep[e.g.,][] {Parker+58,Schatzman+62,Kraft+67,WeberDavis+67,Mestel+68,Skumanich+72}. %For sun-like stars, 
This spindown can be approximated by a power law {for sun-like stars} \citep[e.g.,][]{Dobbs-Dixon+04}, where faster-spinning stars lose more angular momentum per unit time. Over the first gigayear, this dramatically reduces the dispersion in rotational velocities of sun-like stars, and thereafter their spins evolve slower, following the Skumanich relation of $\dot{S}\propto t^{-1/2}$ \citep{Skumanich+72}, where $S$ denotes the stellar spin. {More massive stars spin down differently: }above the ``Kraft Break" at $T_{\rm eff}>6250\,$K \citep{Kraft+67}, stars typically experience little angular momentum loss via magnetic braking due to thin convective zones that cannot drive sufficient wind to achieve angular momentum loss. 

Most old, sun-like stars have similar spin rates \citep[about a few weeks;][]{Delorme+11, Meibom+15,Hartman+10}. The spin of old ($\sim$few Gyrs) stars is not correlated with the young stellar spins, and decreases with time \citep[e.g.,][]{Barnes+03,Skumanich+72}. 
This range of initial conditions and evolution implies that planets are exposed to a wide range of stellar $J_2$ values during their first gigayear. Therefore, the planets are exposed to a time-varying precession again after the disk evaporates. This can produce a sweeping secular resonance similar to those caused by the evolving disk. This $J_2$ sweeping resonance can affect three (or more) planet systems, freezing in excited inclinations and eccentricities \citep{Brefka+21,Faridani+23}, and can even excite two-planet systems {in a resonance} if the inner planet has sufficient angular momentum \citep[e.g.,][]{Spalding+16}. {In these resonances, the planets all torque one another, exchanging angular momentum, and the planet with the least angular momentum more easily has its eccentricity and inclination affected by these angular momentum transfers. If planets have similar masses, then the innermost planet will have the least angular momentum and will be the most excited. Angular momentum conservation dictates that the other planets’ orientation will have to compensate. }

In this work, we focus on observed systems and analyze their secular, planet-planet, and planet-star interaction, with a focus on exploring the potential impact of evolving $J_2$ on their configuration, and whether evolving $J_2$ produces sweeping secular resonances in them. We find that about a fifth of all 3-planet observed transiting systems have configurations that indicate they have undergone sweeping resonances due to evolving $J_2$ in their pasts. This potentially can be used, for example, to constrain the initial conditions of exoplanet systems or to even constrain the initial spin of the host star.

We pedagogically describe the sweeping resonances via the evolving $J_2$ phenomenon and demonstrate its applicability to observed transiting systems. This paper is structured as follows: in Section \ref{sec:math} we construct the secular model used to explore the secular resonances of interest, in Section \ref{sec:initcond} we demonstrate the impact of system initial conditions on whether sweeping resonances occur and their strengths, and in Section \ref{sec:obs} we apply these calculations to $31$ three-planet transiting systems and show approximately one fifth of them may have experienced sweeping resonances in their past. In Section \ref{sec:discussion} we discuss the implications of these findings and what effects might influence the fraction of systems that encounter inclination-driving secular resonances via evolving $J_2$, and in Section \ref{sec:conclusion} we conclude.

\section{Secular Resonances and the Relation to $J_2$}\label{sec:math}

\begin{figure}
    \includegraphics[width=\linewidth]{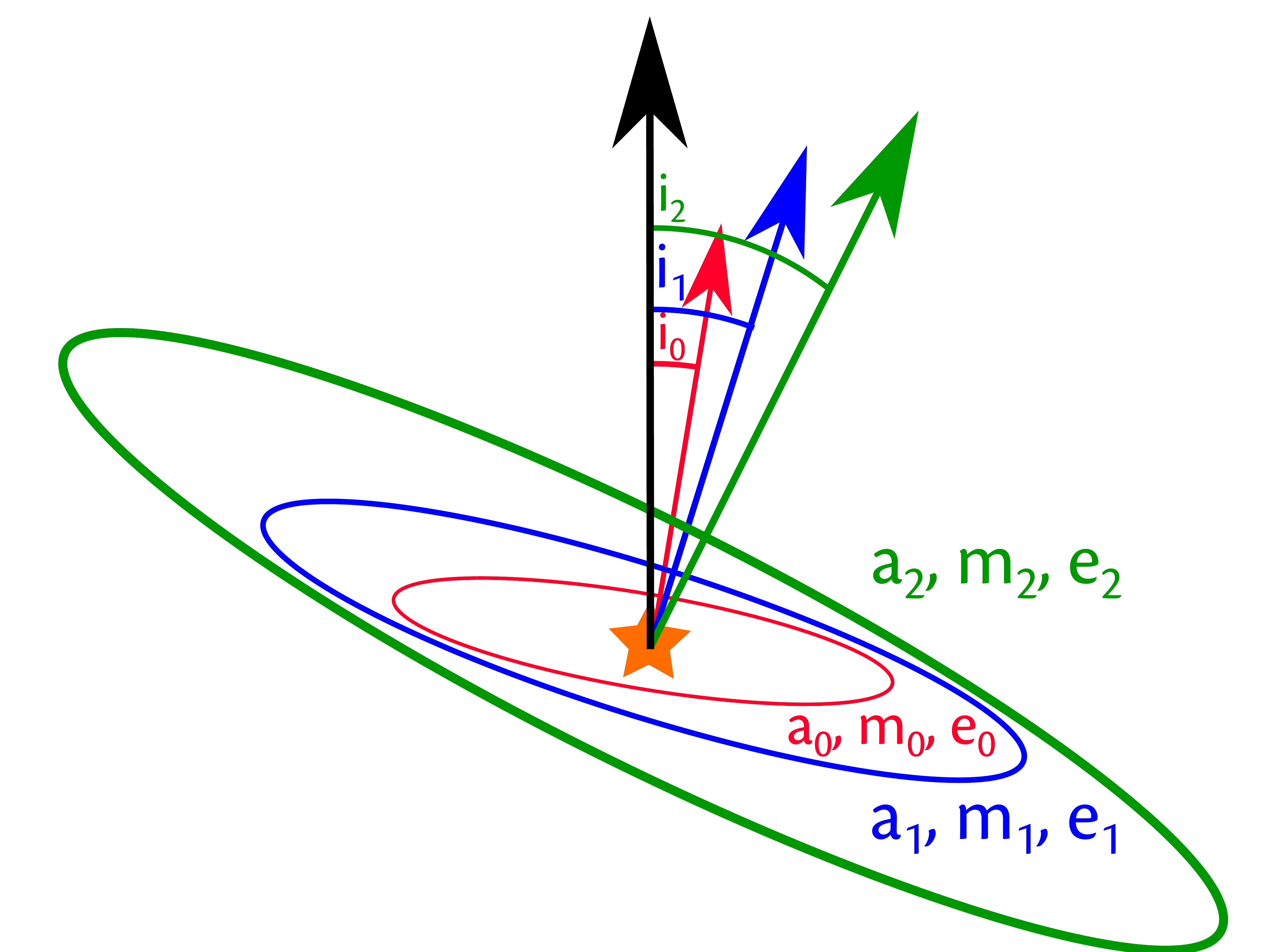}
    \caption{
    A simplified depiction of the three-planet systems considered in this work {(not to scale)}, showing the labeling convention of the planets adopted.
    }
    \label{fig:schematic}
\end{figure}

Consider a star with a mass $M_\star$ and a radius $R_\star$, orbited by a three-planet ($j=0,1,2$) system in which the jth planet has a mass $m_j$,  semi-major axis $a_j$, period $P_j$ (orbital frequency $n_j=2\pi/P_j$), eccentricity $e_j$, longitude of ascending node $\Omega_j$, longitude of periapsis $\varpi_j$, and inclination with respect to the star's spin axis $I_j$. A schematic of the system is depicted in Figure \ref{fig:schematic}. For completeness, here, we specify the Laplace-Lagrange equations. Given the  secular disturbing function $\mathcal{R}^{(\rm sec)}$ \citep[e.g.,][]{Murray+00book}
\begin{equation}\label{eq:R}
\begin{aligned}
\mathcal{R}_j^{(\rm sec)}=& n_j a_j^2\bigg[\frac{1}{2} A_{j j} e_j^2+\frac{1}{2} B_{j j} I_j^2  \\  
&+\sum_{k=1, k \neq j}^N A_{j k} e_j e_k \cos \left(\varpi_j-\varpi_k\right) \\ 
&+\sum_{k=1, k \neq j}^N B_{j k} I_j I_k \cos \left(\Omega_j-\Omega_k\right) \bigg] ,
\end{aligned}    
\end{equation}
where
\begin{equation}
    \begin{aligned}
    A_{jj} =& n_j\bigg[ \frac{3}{2}J_2 \left( \frac{R_\star}{a_j} \right)^2 \\
    &+ \sum_{k=1,k\neq j} \frac{m_k}{4\pi \left( M_\star +m_j\right)} \alpha_{jk} \overline{\alpha}_{jk} f_\psi (\alpha_{jk}) \bigg] \ ,        
    \end{aligned}
    \label{eq:A_matrix_start}
\end{equation}
where $J_2$ is the quadrupole moment of the star. Additionally, 
\begin{equation}
    A_{jk} = -\frac{n_j m_k}{4\pi \left( M_\star +m_j\right)} \alpha_{jk} \overline{\alpha}_{jk} f_{2\psi} (\alpha_{jk})  \ ,
\end{equation}
\begin{equation}
\begin{aligned}
    B_{jj} &= -n_j\bigg[ \frac{3}{2}J_2 \left( \frac{R_\star}{a_j} \right)^2 \\
    &+ \sum_{k=1,k\neq j} \frac{ m_k}{4\pi \left( M_\star +m_j\right)} \alpha_{jk} \overline{\alpha}_{jk} f_\psi (\alpha_{jk}) \bigg] \ ,
    \end{aligned}
\end{equation}\label{eq:bmat}
\begin{equation}
    B_{jk} = \frac{n_j m_k}{4\pi \left( M_\star +m_j\right)} \alpha_{jk} \overline{\alpha}_{jk} f_\psi (\alpha_{jk})  \ ,
    \label{eq:B_matrix}
\end{equation}
where
\begin{equation}
    \alpha_{jk} = \min \left(\frac{a_j}{a_k},\frac{a_k}{a_j}  \right) \ ,
\end{equation}
\begin{equation}
    \overline{\alpha}_{jk} = \min \left(\frac{a_j}{a_k},1  \right) \ ,
\end{equation}
\begin{equation}
    f_{\psi}(\alpha_{jk})=\int_{0}^{2 \pi} \frac{\cos \psi}{\left(1-2\alpha_{jk} \cos \psi+\alpha_{jk}^{2}\right)^{2}} \mathrm{d} \psi\ ,
\end{equation}
and
\begin{equation}
    f_{2 \psi}(\alpha_{jk})=\int_{0}^{2 \pi} \frac{\cos 2 \psi}{\left(1-2\alpha_{jk} \cos \psi+\alpha_{jk}^{2}\right)^{\frac{3}{2}}} \mathrm{d} \psi \ .
    \label{eq:fpsi_end}
\end{equation}
Using the disturbing function, Eq.~\ref{eq:R}, we can find the time evolution of the system's orbital parameters as follows
\begin{equation}\label{eq:LL_orbital_elements}
\begin{aligned}
\dot{e}_j & =-\frac{1}{n_j a_j^2 e_j} \frac{\partial \mathcal{R}_j^{(\rm sec)}}{\partial \varpi_j}, & \dot{\varpi}_j=+\frac{1}{n_j a_j^2 e_j} \frac{\partial \mathcal{R}_j^{(\rm sec)}}{\partial e_j}, \\
\dot{I}_j & =-\frac{1}{n_j a_j^2 I_j} \frac{\partial \mathcal{R}_j^{(\rm sec)}}{\partial \Omega_j}, & \dot{\Omega}_j=+\frac{1}{n_j a_j^2 I_j} \frac{\partial \mathcal{R}_j^{(\rm sec)}}{\partial I_j} ,
\end{aligned}    
\end{equation}

For all planets, a small correction to their $\varpi$ precessional frequency can be made to account for (to first-order) general relativity,
\begin{equation}\label{eq:GR_orbital_elements}
    \dot{\omega}_{j,GR} = \frac{3 (GM_\star)^{3/2}}{c^2 a_j^{5/2} (1-e_j^2)} \ .
\end{equation}
We include this term for all simulations in this work. However, for all the systems of interest, the planets orbit too far out for this term to ever dominate over the planet-planet interactions or the precession caused by $J_2$. In other words, we see that the GR precession term has very little impact in the simulations. Therefore, for our analytic calculations, we omit this term, as it is negligible. Moreover, as GR precession primarily impacts eccentricity evolution (as eccentricity affects the conjugate momentum associated with the $\varpi$ angle), and in this work we are primarily concerned with inclination evolution, the omission of GR is not expected to have significant impact. When eccentricity evolution is of primary concern, GR can have significant impacts \citep[see e.g.,][]{Naoz+12GR,Pu+15,Denham+19,wei+21,Faridani+22,Hansen+20}, and can be incorporated into the Laplace-Lagrange $A$ matrix if desired \citep[see e.g.,][]{Marzari+20}.

Here, we allow the stellar quadrupole moment to evolve as a function of time, adopting the following relation \citep[e.g.,][]{Spalding+16}:
% can be approximated by \citep[e.g.,][]{Spalding+16},
\begin{equation}
    J_2 = \frac{1}{3} k_2 \left(\frac{S}{S_b}\right)^2,
    \label{eq:j2_formula}
\end{equation}
where $k_2$ is the Love number of the star, $S$ represents the star's spin, and $S_b$ is the breakup spin denoted by $S_b = \sqrt{GM_\star/R_\star^3}$.
During the main sequence stage, the stellar spin slows down as it loses angular momentum due to magnetic stellar winds -- a process known as magnetic braking \citep[e.g.,][]{Parker+58,Schatzman+62,WeberDavis+67,Mestel+68}. We adopt the spindown rate following \citet{Dobbs-Dixon+04}, 
\begin{equation}\label{eq:spindown}
    \dot{S} = -\alpha S^3,
\end{equation}
{with $\alpha = 1.5\times10^{-14}$. The effect on $J_2$ is shown in \citet[][Figure 1]{Faridani+23}, duplicated here in Figure \ref{fig:j2evolution}. Briefly, the ranges shown in Figure \ref{fig:j2evolution} correspond to the possible range of initial spins for young sun-like stars, and the evolution is derived from Equation \ref{eq:spindown} and from internal parameters derived from a MESA model \citep[][]{Paxton2011, Paxton2013, Paxton2015, Paxton2018, Paxton2019, Jermyn2023}. {The power-law spin-down relation is chosen for simplicity. The exact functional form of $J_2$ evolution has little effect as long as $\dot{J_2}$ is slower than other relevant timescales in the system \citep[see e.g.,][]{Becker+20,Faridani+23}.}

\begin{figure}
    \includegraphics[width=0.961\linewidth]{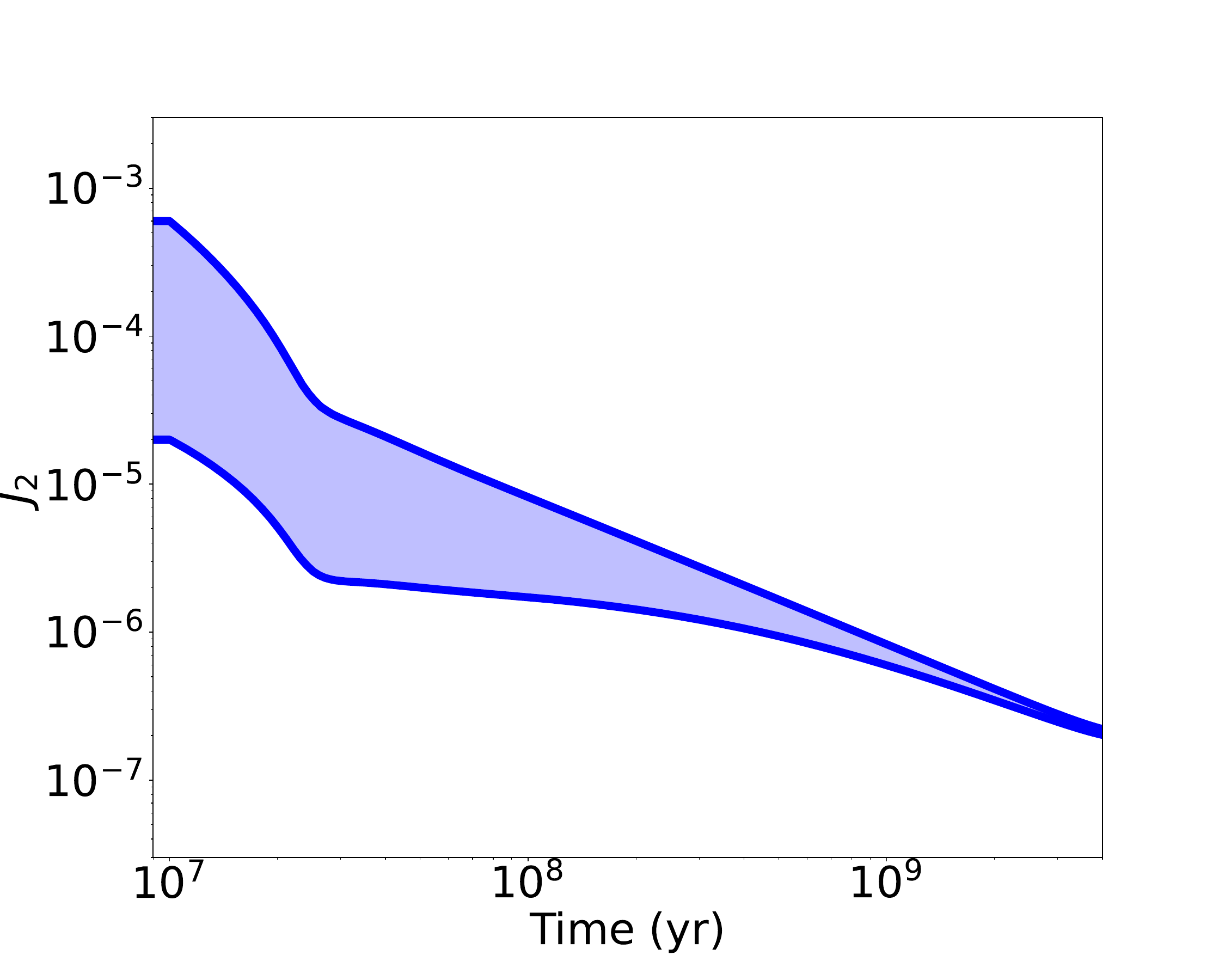}
    \caption{
    Depiction of a range of possible $J_2$ evolutions for sun-like stars. Figure used with permission from \citet{Faridani+23}, and that work describes the models used to generate it.
    }
    \label{fig:j2evolution}
\end{figure}

In \citet{Faridani+23}, we noted that as $J_2$ evolves, an exoplanetary system may be swept into a resonance that leaves an imprint on the planet's eccentricity and inclination. In preparation to map the resonances for a wide range of the planets' parameter space, we first depict an example. Specifically, consider the system Kepler-9 {with parameters given in }Table \ref{tab:Kepler-9}.  The inner planet mass was determined using the mass-radius relation described in \citet{M-RRelationBaron+23}.

In Figure \ref{fig:timeEvolutionKepler9}, we show the time evolution of Kepler-9's planetary system.  Specifically, we show the mutual inclination between the planets (left top panel) and eccentricity (left bottom panel) as well as the alignment or misalignment of the outer two planets' longitude of ascending notes (top right panel) and longitudes of periapsis (bottom right panel). {Overplotted on all panels are the eigenfrequencies of the system (blue and red).} Mutual inclinations were obtained by combining $I_j$ and $\Omega_j$ to get the 3D planetary angular momentum vector for each planet and the mutual inclination was calculated using a dot product. The initial mutual inclination between the outer planets is chosen to be $2.5^\circ$. For simplicity, we adopt that the initial longitudes of ascending node of the planets are set to (approximately) zero. {This evolution was obtained by integrating Equations \ref{eq:LL_orbital_elements} (which include $J_2$ precession) and adding GR precession. As Equations \ref{eq:LL_orbital_elements} are secularly averaged, the evolution in Figure \ref{fig:timeEvolutionKepler9} is therefore purely secular.

% {Note that the transfer of angular momentum as the outer planets and the inner planet are caught in resonance at $t\sim 20\,$Myr occurs rapidly but is not an instantaneous ``kick." It is, rather, a smoother transition as capture into resonance occurs, similar to resonances described in, for example, \citet{Batygin+16Jup}, or \citet{Petrovich+20}. The resonant transition occurs over $\sim 2\times 10^5$ years. This is much slower than the secular eigenfrequencies of the system, which are on the order of $100-1000$ years, though one $\Omega$ eigenfrequency does remain very slow at a timescale of $10^5-10^6$ years.}} 

\begin{figure*}
    \includegraphics[width=\linewidth]{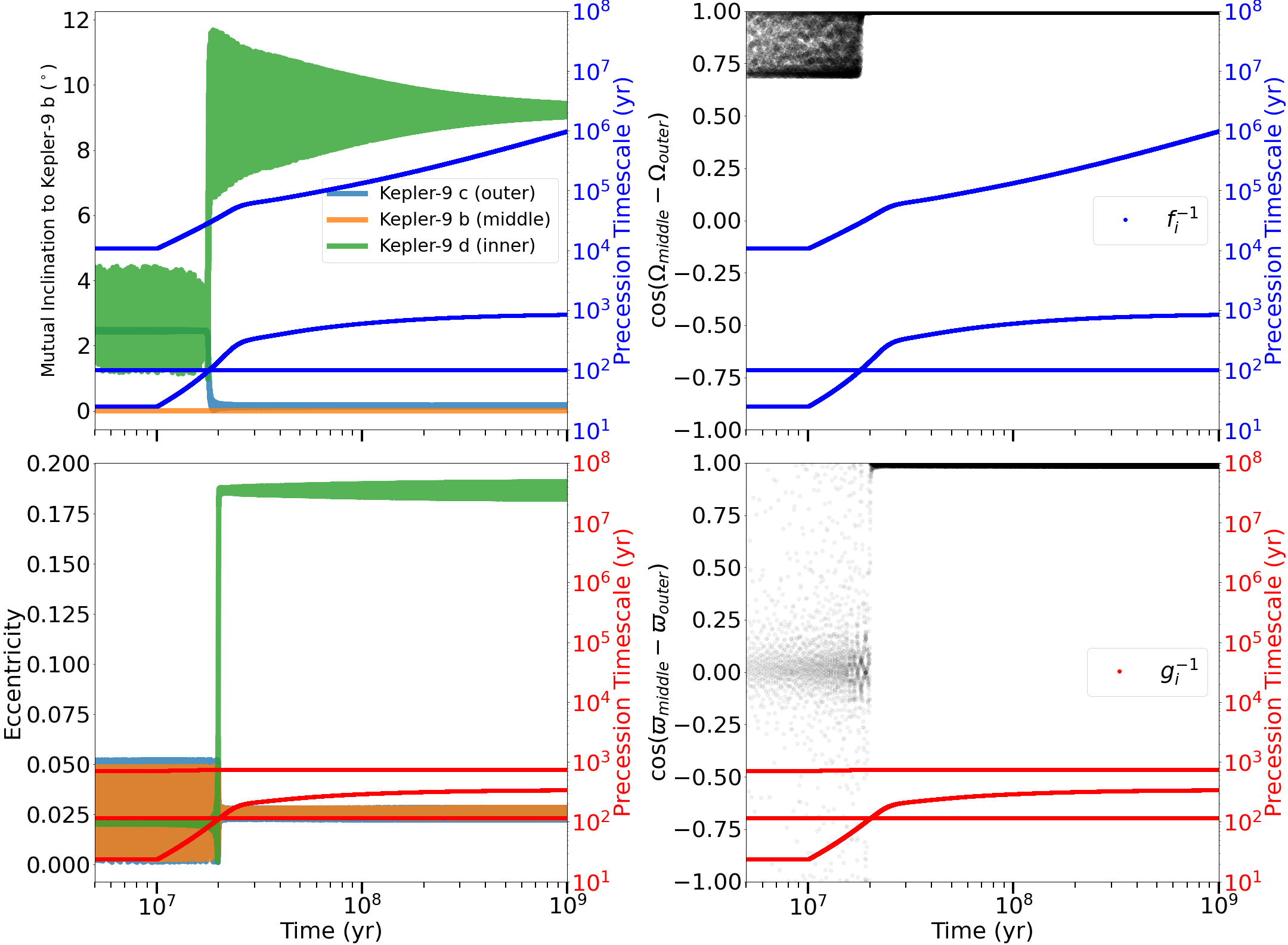}
    \caption{
    Time evolution of sun-like star Kepler-9's planetary system. 
    {Upper Left:} Mutual inclination between planets b, c, and d and Kepler-9 {b (the middle planet)} as a function of time. {Lower Left:} Eccentricity of Kepler 9 b, c, and d as a function of time. {Upper Right:} Evolution of the difference in longitudes of ascending node $(\Omega)$ between Kepler-9 b and c (the middle and outer planets) as a function of time. The alignment near $2\times 10^7\,$years marks the resonance that misaligns Kepler-9 d from the rest of the planets, and simultaneously locks Kepler-9 b and c into coplanarity. {Lower Right:} Evolution of the difference in longitudes of periapsis $(\varpi)$ between Kepler-9 b and c as a function of time. 
    \\
    {{Overplotted on all figures are the Laplace-Lagrange eigenfrequencies $f_i$ and $g_i$.}}
    }
    \label{fig:timeEvolutionKepler9}
\end{figure*}

\begin{table*}
%\tiny
\centering
\begin{tabular}{lcccccccc}
% \hspace{-1cm}
\hline \hline Object & Mass & Radius & $a(\mathrm{au})$ & ${e}^\dag$ & ${i}^\dag$ & ${\Omega}^\dag$ & ${\omega}^\dag$ & Reference \\
\hline Star & $1.09\,M_{\odot}$ & $1.02\,R_{\odot}$ & & & & & & \citet{Torres+11} \\
% USP & $10^{-6}$ & $0.03$ & $10^{-3}$ & $10^{-3} \mathrm{rad}$ & $1$ \\
Kepler-9 d & $\textbf{3.342}\,M_{\oplus}$ & $1.64\,R_\oplus$ & $0.0273$ & $0.021$ & $0.57^\circ$ & $1.61^\circ$ & $89.5^\circ$ & \citet{Torres+11}\\
Kepler-9 b & $43.4\,M_{\oplus}$ & $8.29\,R_\oplus$ & $0.1437$ & $0.026$ & $1.72^\circ$ & $1.58^\circ$ & $123.36^\circ$ & \citet{Borsato+19} \\
Kepler-9 c & $29.9\,M_{\oplus}$ & $8.08\,R_\oplus$ & $0.23015$ & $0.045$ & $4.21^\circ$ & $0.29^\circ$ & $38.47^\circ$ & \citet{Borsato+19} \\
\hline
\end{tabular}
\caption{Initial conditions of Kepler-9 used in Figure \ref{fig:timeEvolutionKepler9}. Bolded masses indicate masses calculated using the mass-radius relation shown in \citet{M-RRelationBaron+23}. Semimajor axes are derived from periods rather than the associated references' reported value to ensure the observed period ratios are maintained.  $^\dag$: parameters randomized according to the scheme described in Table \ref{tab:parametVariationSims}.}
\label{tab:Kepler-9}
\end{table*}

Initially, $ J_2$ is high, and the eccentricities and mutual inclinations oscillate about their initial value and do not evolve significantly for the first few Myr. This is because the $J_2$ evolution in Figure \ref{fig:j2evolution} remains constant for the first $10\,$ Myr.\footnote{see Appendix \ref{sec:timing} for a discussion}. {The inner planet initially has the majority of its eccentricity and inclination oscillation amplitude in the shortest-period eigenmode. The outer two planets have vanishingly small amplitude in this mode. However, as $J_2$ evolves, for both inclination and eccentricity, this mode eventually has the same period as one of the other two modes--bringing the inner planet into resonance with its companions.} Kepler-9 is swept into a resonance, exciting Kepler-9d's eccentricity and inclination, and the outer two planets' longitudes of ascending nodes lock together ($\Omega_b=\Omega_c$). A similar effect occurs in the eccentricities and longitudes of pericenter ($\varpi_b=\varpi_c$), though at a later time in the evolution. The inclination excitation occurs at $t\approx 17\,$Myr and the eccentricity excitation at $\approx\, 22$ Myr.  {Note that despite appearing instantaneous, the exchange of angular momentum that excites the inner planet takes place over $\sim 2\times 10^5$ years. However, despite the exchange occurring over such a long timescale, the exchange is not an adiabatic resonant capture--the inner planet does not remain in resonance with the outer two planets as the colliding eigenvalues diverge. The amplitude of the largest-amplitude eigenmode of the inner planet (i.e., $\max (e_{00},e_{01},e_{02})$ and $\max (I_{00},I_{01},I_{02})$ in Eq \ref{eq:eigencomposition}) is an adiabatic invariant before and after the resonant passage \citep[see e.g.,][for a complete treatment]{Ward+76}, but transitions to a new value during the resonant passage. Moreover, the resonance is only ever achieved by the inner planet's highest-amplitude eigenfrequency approaching another eigenfrequency from above (see the overplotted eigenmode periods in Figure \ref{fig:timeEvolutionKepler9}). To approach from below would require $J_2$ to increase with time.}

Clearly, the effects of resonances mediated by $J_2$ can be drastic. Thus, exploring the maximum inclination (and eccentricity) achieved for a range of $J_2$ values is a promising avenue for evaluating the impact of evolving $J_2$ in different systems. 
For a fixed value of $J_2$ and neglecting General Relativity precession, the maximum value of the planets' inclinations over all time can be calculated as follows\footnote{Including GR does not change the calculation when finding maximum inclinations because GR precesses the argument of pericenter instead of the longitude of ascending node. The maximum eccentricity, however, is affected by GR, and when a planet orbits close enough to its host that it has a period shorter than a few days, this calculation may not be accurate.}. Let $g_i$ and $f_i$ be the eigenvalues of the matrices $A$ and $B$ and $\bar{e}_{ij}$ and $\bar{I}_{ij}$ the matrix of normalized eigenvectors of $A$ and $B$. It can be shown that for $h_j$, $k_j$, $p_j$, and $q_j$ defined as 
\begin{equation}
\begin{aligned}
&h_j=e_j \sin \varpi_j, \quad k_j=e_j \cos \varpi_j,\\
&p_j=I_j \sin \Omega_j, \quad q_j=I_j \cos \Omega_j ,
\end{aligned}
\end{equation}
their time evolution, for a fixed value of $J_2$, is 
\begin{equation}\label{eq:eigencomposition}
\begin{array}{ll}
h_j=\sum_{i=1}^N e_{j i} \sin \left(g_i t+\beta_i\right), & k_j=\sum_{i=1}^N e_{j i} \cos \left(g_i t+\beta_i\right), \\
p_j=\sum_{i=1}^N I_{j i} \sin \left(f_i t+\gamma_i\right), & q_j=\sum_{i=1}^N I_{j i} \cos \left(f_i t+\gamma_i\right),
\end{array}
\end{equation}
for scaled $e_{ji}$ and $I_{ji}$ defined as, $S_i \bar{e}_{j i}=e_{j i}$ and $ T_i \bar{I}_{j i}=I_{j i}$, and phases $\beta_i$ and $\gamma_i$. 
{This time evolution is only true for a fixed value of $J_2$ because changing $J_2$ changes the eigenvalues $g_i$ and $f_i$, as $A_{jj}$ and $B_{jj}$ are functions of $J_2$.}

Therefore, {for a fixed $J_2$,} the maximum eccentricity and inclination of planet $j$ can be calculated by finding the values when all the oscillations are in phase: $e_{j, \rm max} = \sum_{i=1}^N |e_{ji}|$, and $I_{j, \rm max} = \sum_{i=1}^N |I_{ji}|$. To find, $e_{ji}$ and $I_{ji}$ the values of the scaling constants $S_i$, $T_i$ are required. $S_i$, $T_i$ and the phases $\beta_i$ and $\gamma_i$ can be found by solving the linear equations,
\begin{equation}
\begin{aligned}
&e_{ji} {S_i \sin \beta_i}=h_j(t=0) , \quad e_{ji} {S_i \cos \beta_i}=k_j(t=0) ,\\
&I_{ji} {T_i \sin \gamma_i}=p_j(t=0) , \quad I_{ji} {T_i \cos \gamma_i}=q_j(t=0)  .
\end{aligned}
\end{equation}
These calculations enable analytical prediction of the inclination induced by $J_2$ if they are repeated for a wide variety of $J_2$ values.

\section{The Effect of Initial Conditions on Secular Resonances}\label{sec:initcond}

\begin{figure*}
    \includegraphics[width=0.961\linewidth]{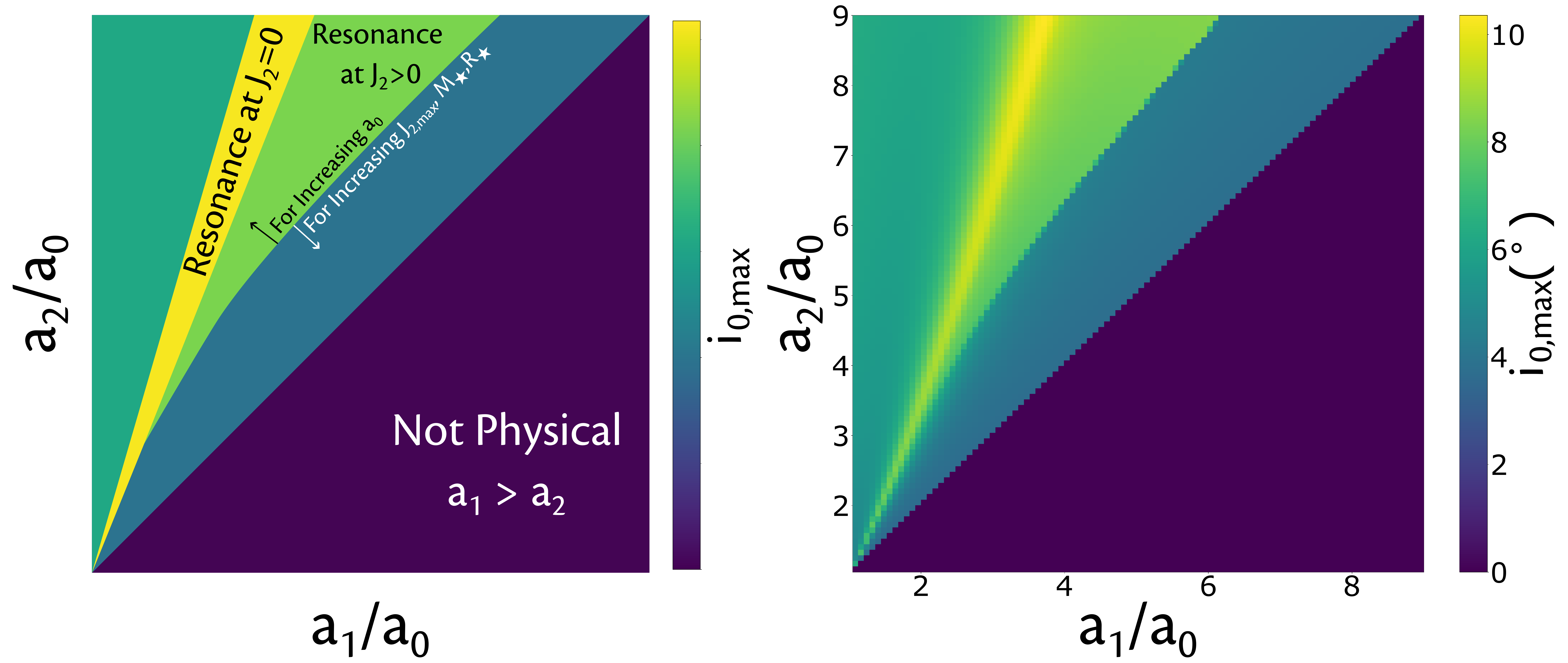}
    \caption{
    Left: Heuristic relationships between companion planet semimajor axis ratios (axes) and maximum inclination induced on the inner planet of a three planet system via planet-planet interactions (color bar) as $J_2$ decreases over time. Annotations describe how variations of system parameters affect the structure. {The boundaries between regions were drawn by eye as a representation of the right panel}.
    \\
    Right: {Analytical calculation of relation between semimajor axis ratios and inner planet maximum inclination over all values of $J_2$ explored by the star as it spins down.}. {The calculations were performed following the prescription described at the end of Section \ref{sec:math}.} Planets 1 and 2 were initialized with a $3^\circ$ mutual inclination. Planet radii were selected as median radii in that orbital position across the $31$ three-planet systems studied in Section \ref{sec:obs} (i.e., the inner planet has the median radius of all inner planets in our sample of 3-planet systems (see Table \ref{tab:allParams}), and so on for middle and outer planet). {The inner planet was set at a semimajor axis of $a_0=0.1\,$au, $J_2$ spanned from $10^{-3}-10^{-8}$, and the inner, middle, and outer planet masses were set to $3.2$, $9.2$, $12.3\,M_\oplus$ respectively using the mass-radius relation of \citet{M-RRelationBaron+23}.}
    }
    \label{fig:smaStrucure}
\end{figure*}

In Figure \ref{fig:smaStrucure}, we show, for a synthetic three-planet system, the relationship between the semimajor axis ratios between the planets and the subsequent inclination induced on the innermost planet by interactions with the outer planets. Planet radii for the inner, middle, and outer planets, were selected as median radii in that orbital position across the $31$ three-planet systems studied in Section \ref{sec:obs}. Planet masses were calculated using the Mass-Radius relation of \citet{M-RRelationBaron+23}. {The analytic calculations described in the previous section describe how to calculate the maximum inclination of the inner planet due to planet-planet interactions and $J_2$ precession for a fixed value of $J_2$. To produce the right panel of Figure \ref{fig:smaStrucure}, these calculations were repeated for many $J_2$ values between $10^{-3}$ and $10^{-8}$, and the maximum inner planet inclination across all $J_2$ values is shown for each semimajor axis ratio in the grid. This calculation produces five visually distinct regions, which we show heuristically in the left panel. The region boundaries in the left panel were drawn by hand.  The notations on the left panel describe how these regions change when parameters are adjusted.}

In the leftmost region, $a_2 >> a_1$, there is no resonance, and inclinations induced on the inner planet are small. To the right of this region is a thin strip where inner planet inclination is maximized. This occurs for only a narrow subset of choices of $a_1$ and $a_2$. This region is caused by secular resonances when $J_2=0$ or when $J_2$ is sufficiently small to be negligible. {These $J_2=0$ resonances occur for the inner planet when the inner planet's oscillation has majority amplitude in the same eigenmode as the outer planets at $J_2=0$, causing it to precess together with one or more of the outer planets. This increases the amount of eccentricity and inclination oscillation of the inner planet relative to the amount of oscillation if the semimajor axis of the inner planet was perturbed slightly \citep[see e.g.,][]{Faridani+22}.}   
% \footnote{{One such $J_2=0$ resonance, named $\nu_6$, forms an edge in the asteroid belt, where asteroids' secular $\varpi$ evolutions become commensurate with Saturn's.}} 
To the right of the $J_2=0$ resonant region is a much larger region of enhanced inclination. This is caused by secular resonances for a range of nonzero, nontrivial $J_2$ values. This region is not in resonance for $J_2=0$, but for every point in the region, there exists some $J_2\neq 0$ which precesses the planets just enough to bring the planets into secular resonance. We call this the ``Sweeping Resonance'' region because for a particular value of $J_2$, only a small portion of the region is in resonance, and it is only as $J_2$ continuously evolves that that resonant portion ``sweeps'' across the parameter space. To the right of this region is a nonresonant region where $a_1 \approx a_2$. To the right of that region, indicated by the dark lower triangle, is a forbidden region by the restriction that $a_2 > a_1$. {Written in the left panel are descriptions of how these regions' boundaries change for different parameter values. 
For increasing $a_0$, 
the sweeping resonant region for $J_2 >0$ decreases in size, but the $J_2=0$ resonance stays in place. For increasing maximum $J_2$, $M_\star$, and $R_\star$, the sweeping resonant region for $J_2 >0$ increases in size.} 

\begin{figure*}
    \includegraphics[width=\linewidth]{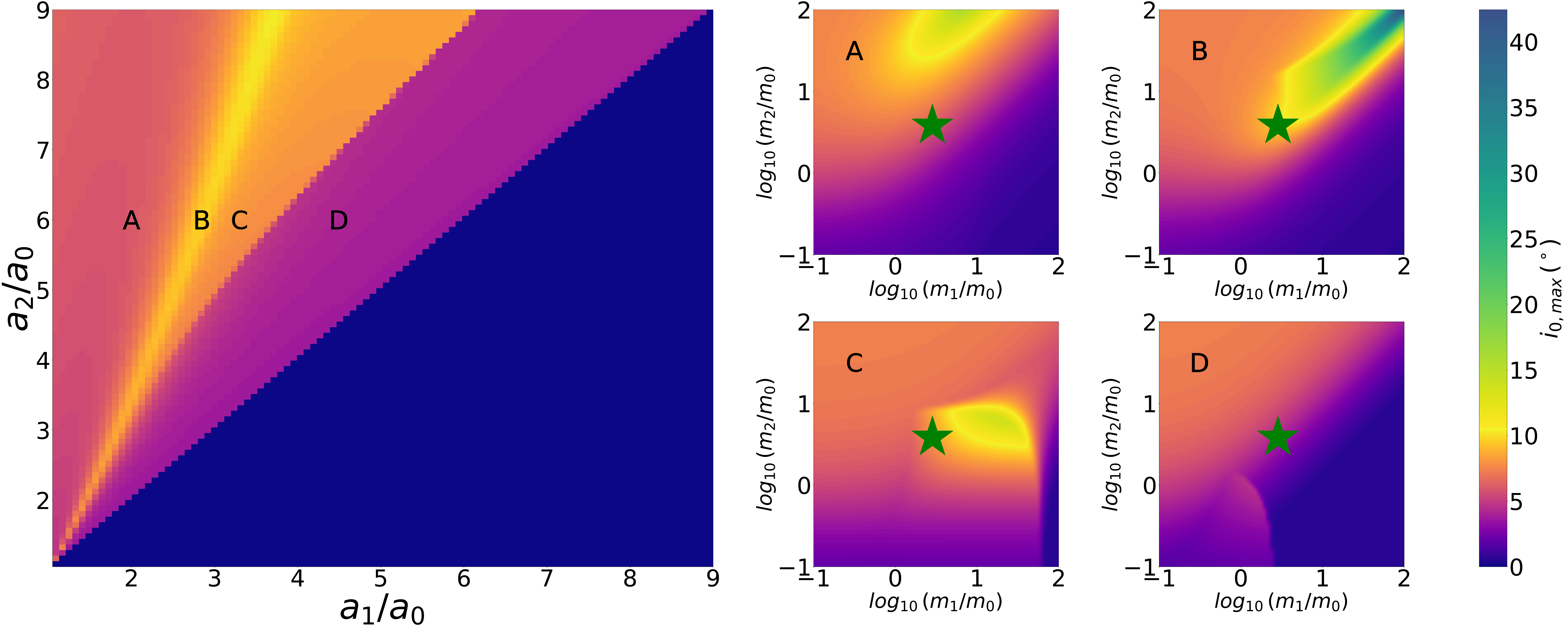}
    \caption{
    Left: Relationship between companion planet semimajor axis ratios (axes) and maximum inclination induced on candidate inner planet via resonance (color bar). Planets 1 and 2 were initialized with a $4^\circ$ mutual inclination
    \\
    Right: Effects of the variation of the masses of the outer two planets at four different sets of semimajor axis ratios shown on the leftmost panel corresponding to the four different qualitative regions shown: {A)} $a_{in}$ and $a_1$ are too close and $a_2$ too far away for the system to enter resonance; {B)} $a_1$ and $a_2$ are such that the inner planet is in resonance with the outer two at $J_2 = 0$; {C)} $a_1$ and $a_2$ are such that the inner planet is in resonance with the outer two for some particular $J_2 \neq 0$; {D)} $a_1$ and $a_2$ are such that the inner planet will never be in resonance with the outer two for any of the $J_2$ values the star will explore. The green stars mark the mass ratio assumed in the left panel.
    }
    \label{fig:massStructure}
\end{figure*}

The general shape shown in the right panel of Figure \ref{fig:smaStrucure} is conserved across many choices of system parameters. Indeed, if $J_2=0$, there are no terms in Eqs (\ref{eq:A_matrix_start})-(\ref{eq:fpsi_end}) that depend on the {individual} semimajor axes of the planets--only the ratios between planets. Moreover, when $J_2=0$, assuming that $m_j << M_\star$ for all planets $j$, then scaling all the planet masses by the same amount will not affect the resulting inclination of the inner planet, as all elements of the matrices will be scaled by the same amount--keeping the eigenvectors constant. However, relaxing $J_2=0$ can turn on the ``Sweeping Resonance'' region. In this case, scaling up all the planet masses reduces the magnitude of the precession caused by $J_2$ compared to planet-planet interactions. This weakening pushes the region where the inner planet experiences ``Sweeping Resonances'' to larger values of $a_1$ and $a_2$, where the mass-boosted planet-planet interactions no longer dominate the $J_2$ precession.

The effect of the planet masses can also be mapped out like we have for the semimajor axes. In Figure \ref{fig:massStructure}, we show the impact of varying the ratios between the outer planets' masses and the inner planet's mass for four different sets of semimajor axes. As a base system, we use the characteristic parameters shown in the {right} panel of Figure \ref{fig:smaStrucure}. The four choices of semimajor axis ratios are labeled on the left panel of Figure \ref{fig:massStructure}, and feature fixed outer and inner planet semimajor axes, with a varying middle planet semimajor axis between them. The middle planet's semimajor axis is varied such that the system explores the four regimes: nonresonant with a small middle planet semimajor axis (labeled `{A}'), static resonance (labeled `{B}'), sweeping resonance (labeled `{C}'), and nonresonant with a large middle planet semimajor axis (labeled `{D}'). The green stars indicate the masses that were used when calculating the left side of Figure \ref{fig:massStructure}. In region `{A},' the nonresonant region, the relationships between mass and inclination induced are smooth, with a slightly larger inclination achieved when $m_2$ increases and $m_1$ stays the same. In region `{B},' the effect of resonance is made clear by the vastly enhanced inclination when both masses $m_1$ and $m_2$ increase relative to $m_{in}$. In region `{C},' the sweeping resonance region, the figure shows the range of masses that allow for sweeping resonance. For $0<\log_{10} (m_1/m_{in})<1.25$ and $-1<\log_{10} (m_2/m_{in})<1$, there is a wedge of parameter space with enhanced inclination. Outside this wedge, the induced inclination is similar to region `{A},' the nonresonant region, indicating that for those masses, no sweeping resonance exists. Region `{D}' appears similar to region `{A}', which is fitting as they are both nonresonant; however, at smaller values of $m_1$ and $m_2$, a similar wedge shape to region `{C}' appears. This indicates that for choices of masses that lie within that wedge, there would be a sweeping resonance at the semimajor axis ratios denoted by region `{D}.'

Notable in Figure \ref{fig:massStructure} is the effect of increasing companion planet masses. In the resonant regions (panels `{B}' and {`C'}), increasing the masses of the outer planets relative to the innermost planet (i.e., moving the green star upward and rightward) tends to increase the amount of misalignment produced by the resonances. However, especially in panel {`C'}, this is not guaranteed. If the masses of the outer two planets are increased past a certain point, then the strength of $J_2$ precession relative to the precession caused by planet-planet interactions decreases, and the requisite $J_2$ to push the inner planet into resonance is higher than the maximum $J_2$ allowed. This defines the top edge of the wedge shape in panel `{C}.' Unlike region `{C}', in region {`B'} where $J_2=0$, there is no maximum outer planet masses beyond which no resonance occurs. The green strip extending to high mass ratios indicates that for a large, given outer planet mass, a middle planet mass exists that brings the inner planet into resonance. {A similar phenomenon is also shown in panel `A'. If, from the mass ratios of the green star, $m_2/m_0$ is increased and $m_1/m_0$ is held constant, then the system can enter the resonant region visible above the green star. In this sense, region `A' is not truly non-resonant for all planet masses.}

\begin{figure*}
    \includegraphics[width=\linewidth]{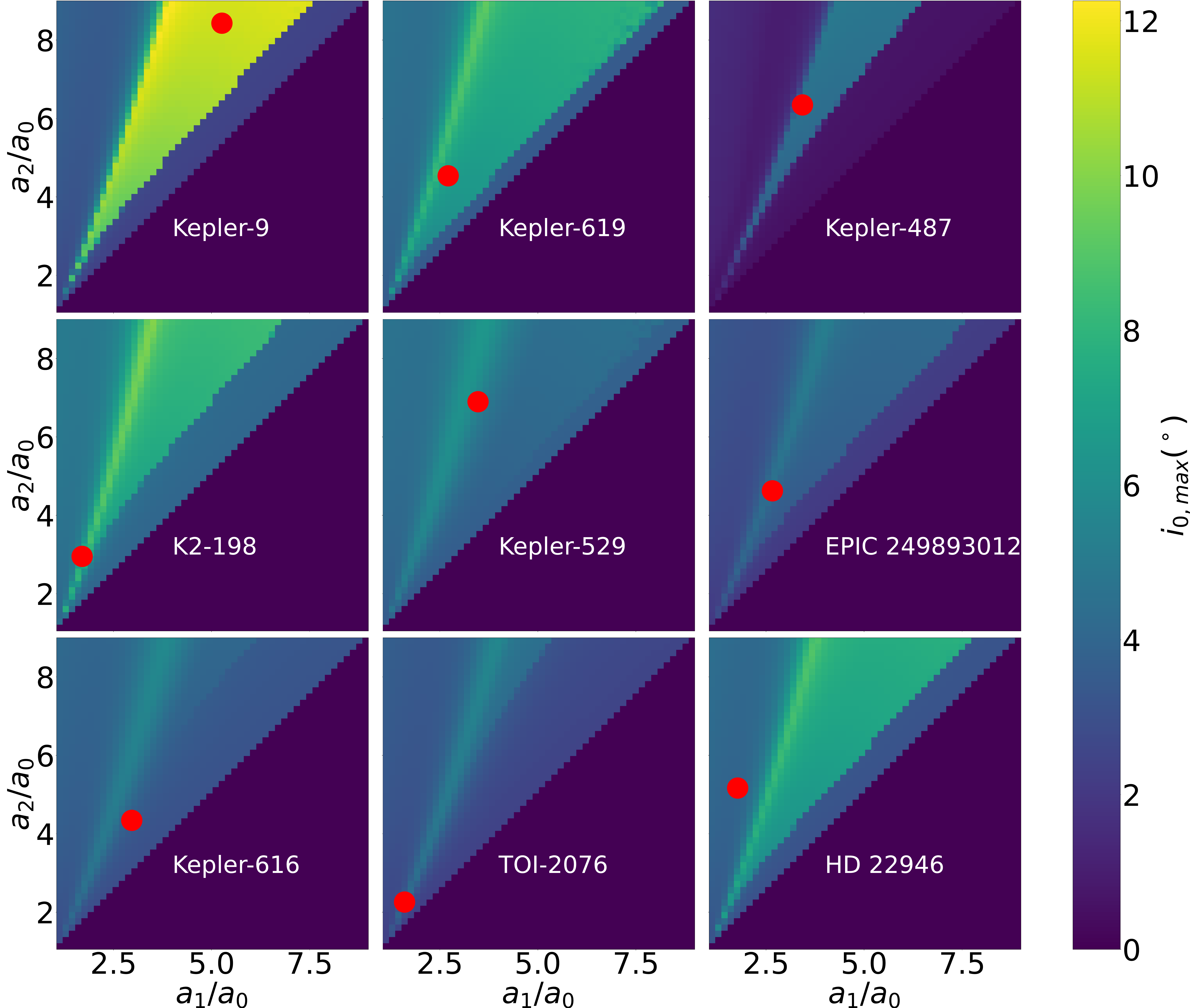}
    \caption{
    Relation between companion semimajor axes and inner planet maxmimum inclination for nine of the $31$ identified three-planet systems in our sample. The initial mutual inclination between the outer planets for each panel was set to $3^\circ$. The top two rows contain only systems that undergo sweeping resonances, and the bottom row contains only systems that do not.
    %Overplotted are curves of mean motion resonances of $3:2,2:1,3:1,4:3,5:2,5:3,5:4$ between each pair of planets.
    }
    \label{fig:9grid}
\end{figure*}

\section{Observed Systems}\label{sec:obs}

Clearly, sweeping secular resonances can have significant impact on exoplanet systems. This naturally poses the question: to what extent are these effects reflected in the known exoplanet sample? To investigate this, we consider 31 known exoplanet systems. These systems were selected from the NASA Exoplanet Archive \citep{exoplanetarchive} with the following criteria: \textbf{1)} The host star has its temperature in the range for FGK-type stars below the Kraft break; \textbf{2)} The host star has its mass and radius measured and is not significantly evolved; \textbf{3)} The planetary system has exactly three observed planets; \textbf{4)} Each planet in the system transits its host star; 
and \textbf{5)} Each planet in the system has its mass or its radius measured. After all {five} criteria have been applied, only $31$ systems remain. In cases where a planet only had its radius measured, its mass was assigned by the Mass-Radius relation given in \citet{M-RRelationBaron+23}. The parameters adopted are listed in Table \ref{tab:allParams}. 

To quantify which systems undergo systematic misalignments similar to those depicted in Figure \ref{fig:timeEvolutionKepler9}, for each of these 31 systems, we run 100 integrations of the secular equations described in Eqs \ref{eq:LL_orbital_elements}. In these integrations, we include the effects of precession due to $J_2$ and GR, and we let $J_2$ evolve according to the upper curve of Figure \ref{fig:j2evolution}. Each simulation was allowed to run for $2\,$Gyr. In each simulation the initial conditions of the planets were randomized according to the prescription given in Table \ref{tab:parametVariationSims}. We find that six of these 31 systems exhibit permanent misalignments between planets mediated by $J_2${, corresponding to about one fifth of the sample}. Similar to the evolution depicted in Figure \ref{fig:timeEvolutionKepler9}, when $J_2$ passes a critical value, the outer two planets in these systems align their orbits. Simultaneously, the innermost planet becomes misaligned {from} the outer two. {The values of $J_2$ and the corresponding times when the misalignment occurs are listed in Table \ref{tab:ResonanceTimes}.}

\begin{table*}
%\tiny
\centering
\begin{tabular}{lcccccc}
% \hspace{-1cm}
\hline \hline Object & Mass, Stellar Radius & $a(\mathrm{au})$ & $e$ & $i$ & $\Omega$ & $\omega$ \\
\hline Star & $\text{From Reference}$ & & & \\
% USP & $10^{-6}$ & $0.03$ & $10^{-3}$ & $10^{-3} \mathrm{rad}$ & $1$ \\
Inner Planet & $\text{From Reference}$ & $\text{From Reference}$ & $\mathscr{R}(0.049)$ & $1^\circ$ & $\mathscr{U}(2\pi)$ & $\mathscr{U}(2\pi)$\\
Middle Planet & $\text{From Reference}$ & $\text{From Reference}$ & $\mathscr{R}(0.049)$ & $\mathscr{R}(0.049 \,\text{rad})$ & $\mathscr{U}(2\pi)$ & $\mathscr{U}(2\pi)$\\
Outer Planet & $\text{From Reference}$ & $\text{From Reference}$ & $\mathscr{R}(0.049)$ & $\mathscr{R}(0.049 \,\text{rad})$ & $\mathscr{U}(2\pi)$ & $\mathscr{U}(2\pi)$ \\
\hline
\end{tabular}
\caption{Randomization Scheme for Simulations. {Simulations were integrated for $2\,$Gyr.} $\mathscr{R}(\sigma)$ represents a random variable drawn from a Rayleigh distribution with scale $\sigma$. $\mathscr{U}(X)$ represents a random variable drawn from a uniform distribution between $0$ and $X$.}
\label{tab:parametVariationSims}
\end{table*}

\begin{table}
%\tiny
% \centering
\begin{tabular}{lcc}
% \hspace{-1cm}
\hline \hline System &  Resonant $J_2$  & $t_{{res}}$ (yr) \\
 Kepler-9     & $1.3\times 10^{-4}$ & $1.7\times 10^7$ \\
 Kepler-619     & $7\times 10^{-5}$ & $2\times 10^7$ \\
 Kepler-487     & $1.1\times 10^{-4}$ & $1.85\times 10^7$ \\
 K2-198     &  {$ {8.3}\times  {10}^{ {-5}}$} &  {$ {2}\times  {10}^ {7}$} \\
 Kepler-529     & $1.8\times 10^{-5}$ & $4.6\times 10^7$ \\
 EPIC 249893012     &  {$ {2.1}\times  {10}^{ {-4}}$} &  {$ {1.5}\times  {10}^ {7}$} \\
\end{tabular}
\caption{Values of $J_2$ for system inclination resonance for the six sweeping-resonant systems identified in Section \ref{sec:obs}. Additionally, the corresponding times for this $J_2$ value are listed (following the top curve of Figure \ref{fig:j2evolution}). 
}
\label{tab:ResonanceTimes}
\end{table}

In Figure \ref{fig:9grid}, we show the structure {of increases of inner-planet inclination} in $a_1/a_0$--$a_2/a_0$ space of nine systems. Each panel of Figure \ref{fig:9grid} represents a different observed system, with the observed values of the semimajor axis ratios (derived from the extremely accurately measured period ratios) marked with a red dot. The top two rows of Figure \ref{fig:9grid} are the six systems from our $31$-system sample which were found in the Monte Carlo simulations to undergo permanent misalignments. 
The bottom row of Figure \ref{fig:9grid} contains three of the $25$ systems whose simulations {did not undergo permanent misalignments. }
The bottom row systems are chosen from various regions of non-sweeping parameter space: high $a_1/a_0$ (Kepler-616; see Table \ref{tab:allParams}), ``static" resonance that does not {sweep, and therefore does not} produce misalignments (TOI-2076; discussed in detail in Section \ref{sec:nonsweeping}), and low $a_1/a_0$ (HD 22946; see Table \ref{tab:allParams}). {The values of $i_{0,max}$ shown in Figure \ref{fig:9grid} are calculated using the method described at the end of Section \ref{sec:math} for initial mutual inclinations between middle and outer planets of $3^\circ$. The predictions of inner-planet inclination of Figure \ref{fig:9grid} and the simulations shown later in the red circles of Figure \ref{fig:alignmentchange} demonstrate good agreement between the analytic prediction and the simulated result.}

{To quantify the misalignment present in a system caused by changing $J_2$, we define the parameter}
\begin{equation}\label{eq:deltaI}
    \Delta i_{jk}^{mut} =\langle {i_{jk}^{mut}}\rangle_{late} -\langle {i_{jk}^{mut}}\rangle_{early} \,\,\,,
\end{equation}
where $\langle {i_{jk}^{mut}}\rangle_{early}$ is the time-averaged mutual inclination between planets $j$ and $k$ before $t=1\,$Myr (this is before $J_2$ evolution is turned on at $t=10\,$ Myr) and $\langle {i_{jk}^{mut}} \rangle_{late}$ is the average mutual inclination between planets $j$ and $k$ during the last $40\,$ Myr of the simulation. A positive value of $\Delta i_{jk}^{mut}$ implies that planets $j$ and $k$ became more misaligned over the course of the simulation, and a negative value implies that planets $j$ and $k$ aligned themselves. We choose to define $\Delta i_{jk}^{mut}$ in terms of time-averages at early and late times to remove dependence on the {initial} phase of {inclination} oscillation. 

In each panel of Figure \ref{fig:alignmentchange}, we show the relationship between the initial mutual inclination between the outer two planets {(x-axis)} and the {change in planets' relative alignment (}y-axes, {$\Delta i_{01}^{mut}$ in red and $\Delta i_{12}^{mut}$ in blue).} The top two rows are systems whose positions in $a_1/a_0$--$a_2/a_0$ space indicate they undergo sweeping resonances. For these systems, there is an approximately linear relationship between initial mutual inclination between outer planets, and the (mis-)alignments between the planets after $J_2$ evolution has had its effect.  {In each panel, the dotted red line represents the minimum mutual inclination between the middle and inner planets such that the inner planet may not co-transit with the middle planet under the simplifying assumption that the middle planet's transit impact parameter $b_1=0$. Many of the sweeping resonant systems would require very large initial mutual inclinations between the outer two planets to make the inner planet no longer transit.} 

{However, that is not the only constraint we can place on the initial conditions. If the outer two planets are prepared with high initial mutual inclinations, under what conditions do they align enough over the course of the simulation such that at the end, they both transit? This is shown in the blue dotted lines. For blue triangles above the blue dotted line, the initial mutual inclination between the outer two planets is {high enough that they may not} transit together (assuming the middle planet's impact parameter $b_1=0$), even after the sweeping resonance from evolving $J_2$ partially aligns their orbits. { Of course, if the longitudes of ascending nodes are favorable, even two greatly mutually inclined planets may still co-transit. Therefore, even systems with red dots above the red dotted line or blue triangles above the blue dotted line may have all three planets transit depending on nodal positioning.} {To account for these two constraints on $i_{mut,12,init}$, regions on Figure \ref{fig:alignmentchange} are shaded. The shaded regions correspond to the ranges of $i_{mut,12,init}$ such that all three planets are guaranteed to co-transit under the $b_1=0$ assumption. The color of the shaded region indicates which planet escapes transit first as $i_{mut,12,init}$ increases--red for the inner planet and blue for the outer planet.
% The region is colored red if the first planet to escape guaranteed co-transit above this $i_{mut,12,init}$ range is the inner planet, and blue if it is the outer planet.
} {Therefore, to be guaranteed consistency with observations, $i_{mut,12,init}$ must be smaller than the rightmost edge of the shaded region.} {For example, Kepler-487 (which experiences sweeping resonance), has only a small range of $i_{mut,12,init}$ values where co-transit is guaranteed. The outer planets of Kepler-487 do not align sufficiently during resonance to overcome large initial mutual inclinations, and at $i_{mut,12,init}\sim 2^\circ$ the outer two planets no longer always co-transit. Systems that do not undergo sweeping resonance--such as Kepler-616, for example--have far stricter restraints on $i_{mut,12,init}$, as they are subject to the geometric requirements for co-transit without a sweeping resonance to align the outer planets.} {Additionally, Kepler-487, as well as K2-198 and EPIC 249893012, also has an interesting property where if $i_{mut,12,init}$ is near zero, then the inner and middle planets will be the ones aligning, and the outer planet misaligns from the rest. This is represented by negative (positive) values of $\Delta i_{01}^{mut}$ ($\Delta i_{12}^{mut}$).}

% For example, even though the inner planet of K2-198 still transits with the middle planet even for high $i_{mut,12,init}$, {the shaded region is small, as }the outer two planets do not align enough from sweeping resonances to {guarantee co-transit}. On the other hand, even if $i_{mut,12,init}$ is set to $10^\circ$ in Kepler-9 or Kepler-619, the outer planets always align enough from the sweeping resonance to co-transit. } {Kepler-487, however, has only a small range of $i_{mut,12,init}$ values where the blue points lie beneath that blue curve and the red points below the red curve. Therefore, only for a narrow range of $i_{mut,12,init}$ are all three planets aligned enough to guarantee a co-transiting configuration. Moreover, Kepler-487 also has an interesting property where if $i_{mut,12,init}$ is near zero, then the inner and middle planets will be the ones aligning, and the outer planet misaligns from the rest. }

{With these constraints on initial inclinations in mind, it may be worthwhile to consider whether they are consistent with known constraints on the exoplanet population. For example, \citet{He+20} and \citet{Millholland+21} find that the \textit{Kepler} sample's properties are consistent with each system having the maximum stable angular momentum deficit \citep[AMD; see e.g.,][]{Laskar+17} distributed randomly, weighted by mass, between the planets. The sweeping resonances described here conserve angular momentum (and therefore AMD), meaning that these resonances do not produce a systematic signal contrary to the maximum-AMD model. However, as can been seen in Figures \ref{fig:timeEvolutionKepler9} and \ref{fig:alignmentchange}, these resonances cause the outer two planets to align and circularize, dumping much of their AMD into the (usually less-massive) inner planet to make its orbit misaligned and eccentric. If the initial distribution of AMD between planets is as assumed by \citet{He+20}, where the more massive planets hold more of the system's AMD, then observations of lower-mass inner planets holding outsized proportions of AMD could be an indication that these sweeping resonances have occurred.}

\begin{figure*}
    \includegraphics[width=\linewidth]{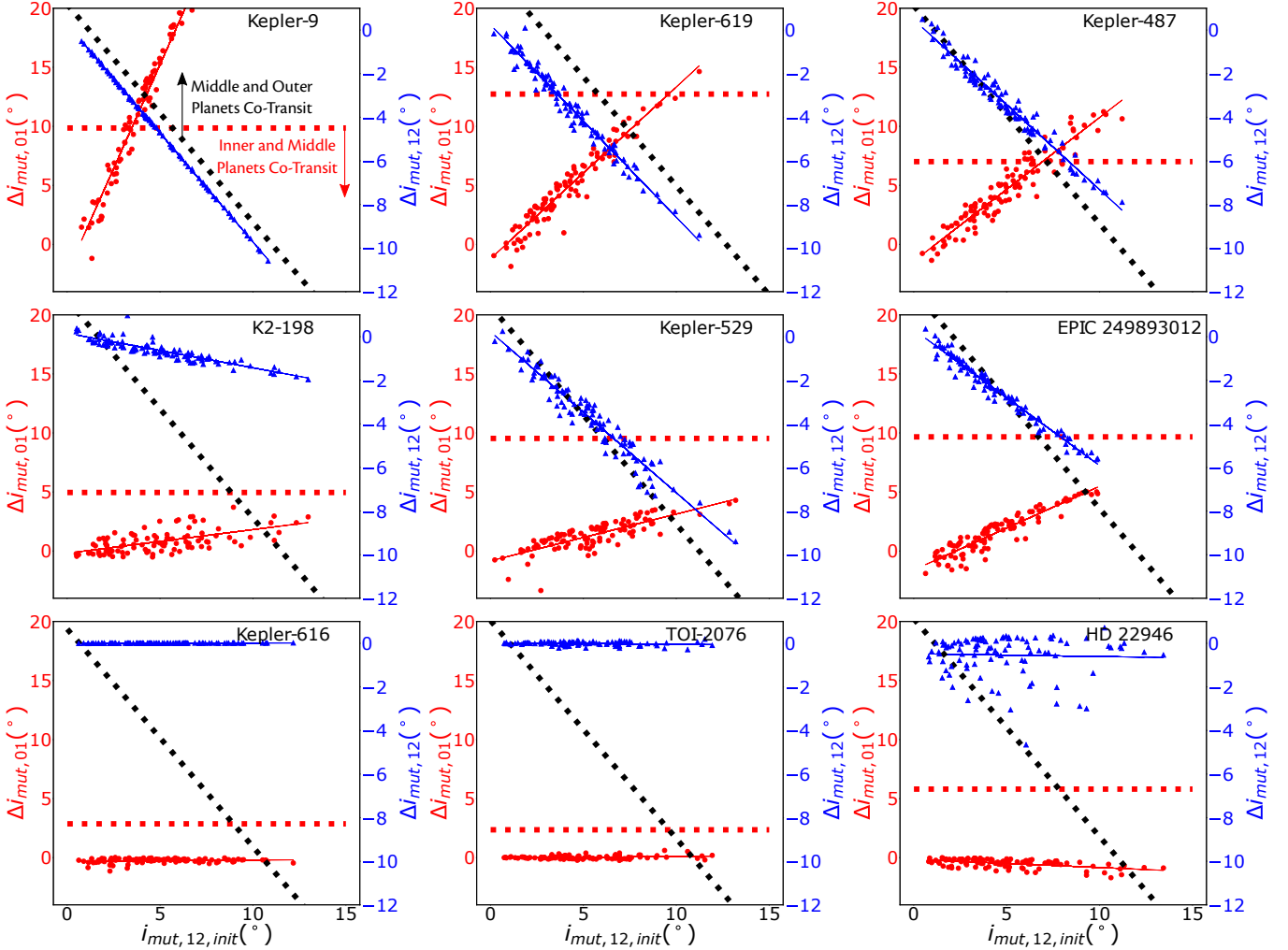}
    \caption{
    Relation between initial mutual inclination between the two outermost planets and changes in alignment between planets over the course of the simulation for nine systems. The top six systems undergo resonance-mediated misalignments. The bottom three do not. In red circles are, for each simulation, the change in alignment between the inner and middle planets, {($\Delta i_{mut,01}$)}--the average mutual inclination at late times between the inner and middle planet minus the initial mutual inclination between the inner and middle planet {(see Eq \ref{eq:deltaI})}. The red dotted line represents the minimum $\Delta i_{mut,01}$ such that the inner planet fails to transit {at the end of the simulation} (under the simplifying assumption that the middle planet has a transit impact parameter $b_1=0$). In blue triangles are, for each simulation, the change in alignment between the middle and outer planets, {($\Delta i_{mut,12}$)}--the average mutual inclination at late times between the middle and outer planet minus the initial mutual inclination between the middle and outer planet. {The blue dotted line represents the maximum value of $\Delta i_{mut,12}$ as a function of $i_{mut,12,init}$ such that the outer two planets {are guaranteed to co-}transit at the end of the simulation if $b_1=0$.} {Lines of best fit are plotted for both the red circles and blue triangles. The shaded regions represent the range of $i_{mut,12,init}$ such that all three planets are guaranteed to transit (again, if $b_1=0$). The color of the shaded region indicates which planet escapes transit first as $i_{mut,12,init}$ increases--red for the inner planet and blue for the outer planet.} The scheme by which orbital parameters were varied between simulations is described in Table \ref{tab:parametVariationSims}.
    % Lines of best fit are shown in their respective color, and their values are listed in Table \ref{tab:lineSlopes}. 
    }
    \label{fig:alignmentchange}
\end{figure*}

\subsection{Sweeping Resonant Systems}\label{sec:sweeping}

In Figure \ref{fig:9grid} we showed six systems that lie in the range of parameter space where sweeping resonances via $J_2$ evolution can occur. Parameters adopted for all $31$ systems in the sample are shown in Table \ref{tab:allParams}. Below we comment on Kepler-9 and EPIC 249893012, as their properties are of note.

\subsubsection{Kepler-9}
Kepler-9 is a well-studied system, discovered by \citet{Holman+10}. As it was one of the first-validated Kepler systems, its parameters have been studied and adjusted many times through transits, transit timing variations, and radial velocity measurements \citep[e.g.,][]{Borsato+19, Albrecht+21, Berger+18,Wang+18,Hadden+17,Holczer+16,Morton+16,Borsato+14,hadden+14,Torres+11,Weiss+24}. The outer two planets are near, but outside, a $2:1$ mean-motion resonance that integrations of the most recent fits indicate is an unstable resonance \citep[e.g.,][]{Antoniadou+20}. The literature suggests that the outer two planets have aligned orbital axes \citep[e.g.,][]{Freudenthal+18}, which we note is consistent with the damping out of any primordial orbital misalignments between them as shown in Figures \ref{fig:timeEvolutionKepler9} and \ref{fig:alignmentchange}.

\subsubsection{EPIC 249893012}
EPIC 249893012 is a sun-like star just beginning its evolution off the main sequence \citep{Hidalgo+20}. As it is no longer a main-sequence star, it would normally have been excluded from this sample. However, as there is substantial evidence this star is just leaving the main sequence, its mass while its $J_2$ was spinning down would have been approximately the same as its current mass and therefore the planets should not have experienced stellar evolution-based migration. Therefore, it is a valid inclusion, and for the purposes of this work, we have set its radius to $1.1\,R_\odot$. 

\citet{Gupta+21} calculates that the inner planet, EPIC 249893012 b, is losing mass at a rate of approximately $10^8\,g/s$. However, if this mass loss rate has been sustained for the entire age of the system $\sim 9\,\mathrm{Gyr}$, the mass of EPIC 249893012 b would have changed by only $0.005\,M_\oplus$ since its formation. Therefore, we adopted EPIC 249893012 b's currently measured mass for use in our simulations, as its atmospheric mass loss does not alter significantly its total mass.

\subsection{The Impact of Static vs Sweeping Resonances}\label{sec:nonsweeping}

As depicted in Figure \ref{fig:smaStrucure}, we identify a region in resonance associated with $J_2=0$ (the yellow strip, labeled ``Resonance at $J_2=0$''). 
Not all of this $J_2=0$ resonant region produces systematic misalignments similar to those demonstrated in Figure \ref{fig:timeEvolutionKepler9}. At small values of $a_1/a_0$ and $a_2/a_0$ or when $a_0$ is large, or both, the precession rate caused by $J_2$ is slow (even at $J_2=10^{-3}$) compared to the precession caused by planet-planet interactions. In this case, $J_2$ precession is not fast enough to shift the location of the secular resonance for any value of $J_2$ the star can be reasonably expected to achieve. We call these unmoving resonances ``static". 

Static resonances notably do not produce the misalignment from evolving $J_2$ that sweeping resonances do. For example, the system TOI-2076 \citep[observed by e.g.,][]{Osborn+22,Hedges+21} lies on such a static resonance as depicted in the bottom middle panel of Figure \ref{fig:9grid} . In Figure \ref{fig:TOI2076Param}, we show the semimajor axis parameter space of TOI-2076 for $J_2=10^{-3}$ (left panel) and $J_2=0$ (right panel). The location of TOI-2076's observed semimajor axis ratios is marked with a red dot. In both panels, TOI-2076 lies on a secular resonance with the same expected inner planet maximum inclination. This indicates that there is no expected change to the system as $J_2$ evolves from $\sim 10^{-3}$ to $0$. Therefore, we do not expect the inner planet to systematically and/or permanently misalign from its companions any more than is provided in its initial conditions. {Performing the eigendecomposition described in Section \ref{sec:math}, calculating the amplitudes of each of the planets in each of the eigenfrequencies of the $B$ matrix, one finds that all three planets have their maximum amplitude in the same mode for any of the values of $J_2$ taken on in Figure \ref{fig:j2evolution}. This mode's period can be seen to vary from $10^4$ years to $10^7$ years in the top row of Figure \ref{fig:timeEvolutionTOI2076}. Therefore, the three planets have their $\Omega$ values precess together forever, with all the angles $\Omega_1-\Omega_0$, $\Omega_2-\Omega_0$, and $\Omega_2-\Omega_1$ librating.}

Indeed, in Figure \ref{fig:timeEvolutionTOI2076}, we show the time evolution of one of the simulations of TOI-2076 from the Monte Carlo. In the {upper right} panel, we can see that the longitudes of ascending nodes between the outer two planets are locked together; however, the upper left panel shows that mutual inclinations do not significantly change. The planets do not misalign. In the bottom middle panel of Figure \ref{fig:alignmentchange}, we can see that even for large initial values of $i_{mut,12}$, no permanent misalignment occurs between the planets in this system--they retain whatever mutual inclinations they are initialized with.

Therefore, the misalignments shown in the sweeping resonant systems depicted in the top two rows of Figure \ref{fig:alignmentchange} arise because these systems are initially not in resonance for high $J_2$ values, then enter resonance as $J_2$ decreases, {increasing the inner planet's inclination predicted from Laplace-Lagrange}. Depending on their locations in $a_1/a_0$--$a_2/a_0$ space, {the planets may} later leave resonance{, but the mark of having passed through resonance is left on their mutual inclinations}. 
{It is the change/increase in predicted inclinations from Laplace-Lagrange due to evolving $J_2$ that produces the misalignments.}

\begin{figure*}
\includegraphics[width=\linewidth]{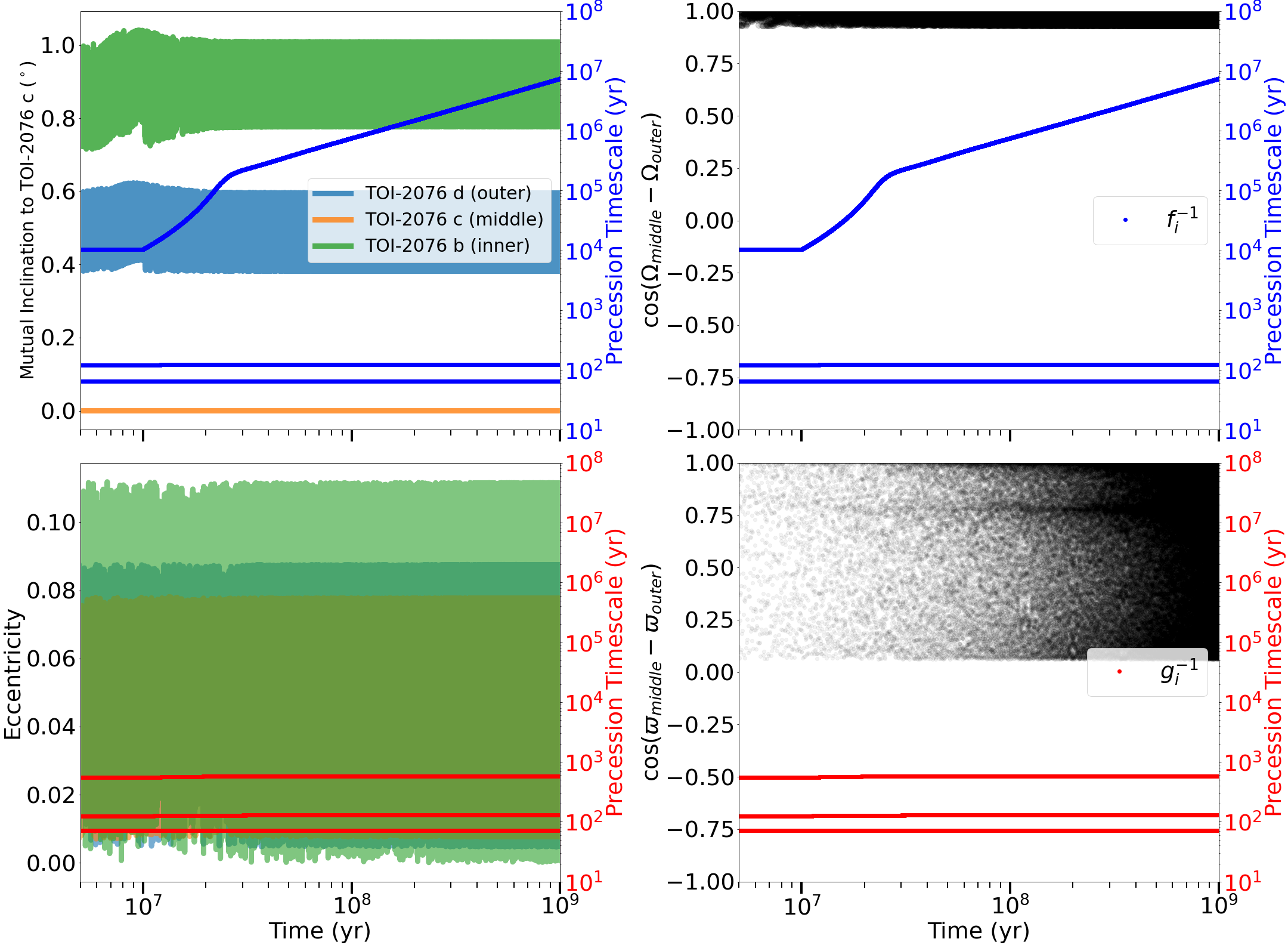}
    \caption{
    Time evolution of K-star TOI-2076's planetary system. 
    {Upper Left:} Mutual inclination between planets b, c, and d and TOI-2076 c as a function of time. {Lower Left:} Eccentricity of TOI-2076 b, c, and d as a function of time. {Upper Right:} Evolution of the difference in longitudes of ascending node $(\Omega)$ between TOI-2076 c and d (the middle and outer planets) as a function of time. {Lower Right:} Evolution of the difference in longitudes of periapsis $(\varpi)$ between TOI-2076 c and d as a function of time. 
    \\
    {{Overplotted on all figures are the Laplace-Lagrange eigenfrequencies $f_i$ and $g_i$.}}
    }
    \label{fig:timeEvolutionTOI2076}
\end{figure*}

\begin{table*}
%\tiny
\centering
\begin{tabular}{lcccccccc}
% \hspace{-1cm}
\hline \hline Object & Mass                         & Radius            & $a(\mathrm{au})$ & ${e}^\dag$ & ${i}^\dag$   & ${\Omega}^\dag$ & ${\omega}^\dag$ & Reference \\
\hline TOI-2076      & $0.82\,M_{\odot}$            & $0.77\,R_{\odot}$     & & & & &                                                                      & \citet{Osborn+22} \\
% USP & $10^{-6}$ & $0.03$ & $10^{-3}$ & $10^{-3} \mathrm{rad}$ & $1$ \\
TOI-2076 b           & $\textbf{10.4}\,M_{\oplus}$ & $2.51\,R_\oplus$   & $\textbf{0.087}$          & $0.032$    & $0.57^\circ$ & $1.25^\circ$    & $107.5^\circ$    & \citet{Osborn+22}\\
TOI-2076 c           & $\textbf{17.2}\,M_{\oplus}$ & $3.50\,R_\oplus$   & $\textbf{0.139}$          & $0.066$    & $1.61^\circ$ & $1.24^\circ$    & $281.4^\circ$  & \citet{Osborn+22} \\
TOI-2076 d           & $\textbf{15.2}\,M_{\oplus}$ & $3.23\,R_\oplus$   & $\textbf{0.196}$          & $0.057$    & $2.03^\circ$ & $1.7^\circ$    & $318.2^\circ$   & \citet{Osborn+22} \\
\hline
\end{tabular}
\caption{Initial conditions of TOI-2076 used in Figure \ref{fig:timeEvolutionTOI2076}. Bolded masses indicate masses calculated using the mass-radius relation shown in \citet{M-RRelationBaron+23}. Bolded semimajor axes indicate semimajor axes calculated from stellar mass and planetary orbital period.  $^\dag$: parameters randomized according to the scheme described in Table \ref{tab:parametVariationSims}.}
\label{tab:TOI-2076}
\end{table*}

\begin{figure*}
    \includegraphics[width=\linewidth]{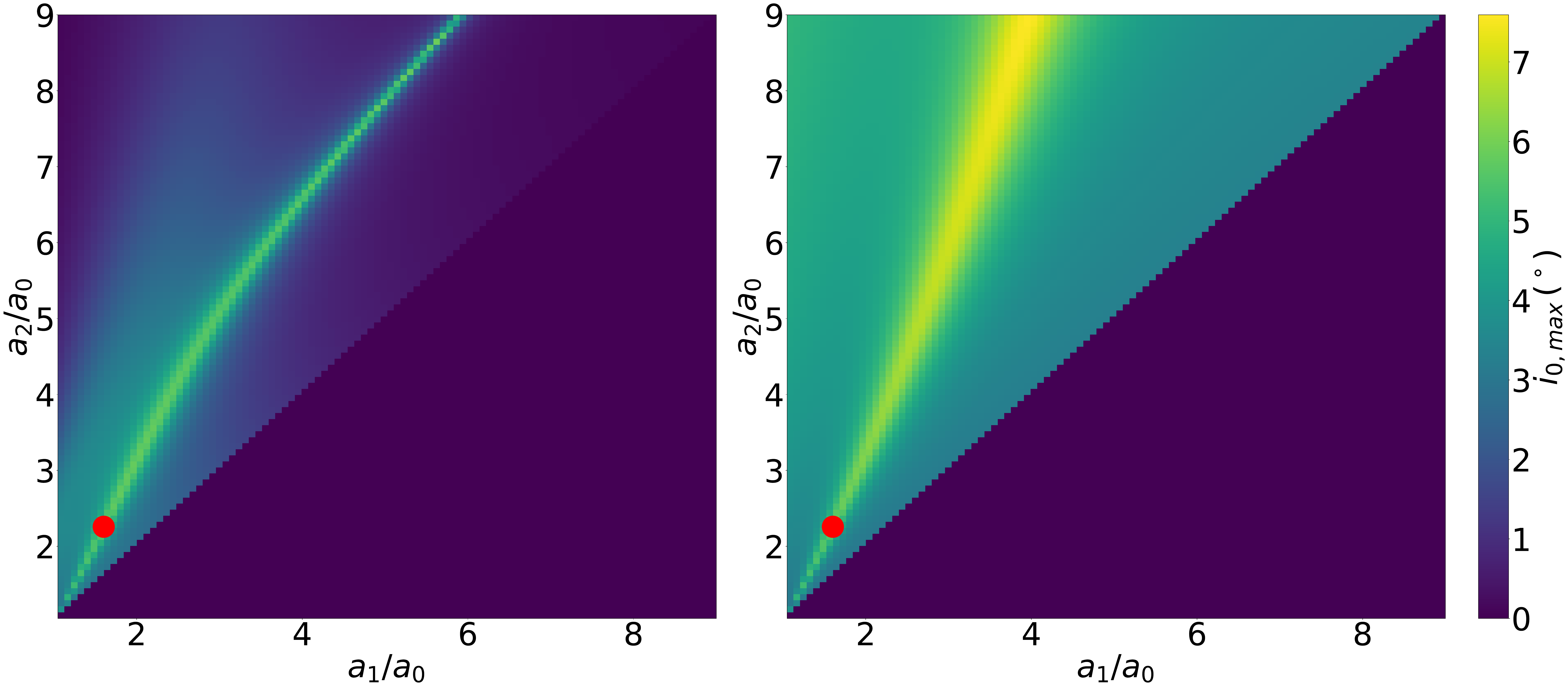}
    \caption{
    Parameter space of TOI-2076 with $J_2$ fixed at $10^{-3}$ (left panel) and $0$ (right panel). The observed locations of the outer planets are shown by a red dot. The mutual inclination between the outer two planets of TOI-2076 was set to $3^\circ$.
    }
    \label{fig:TOI2076Param}
\end{figure*}

\section{Discussion}\label{sec:discussion}
Out of our clean sample of $31$ three-planet transiting systems, we found six systems whose parameters indicate they may have undergone sweeping resonance during their first Gyr due to $J_2$ evolution. 
This corresponds to one system in five, or $\sim 20\%$, indicating that the architectures that produce these misalignments may be common. While performing a full completeness analysis to estimate the true number of systems with these architectures is beyond the scope of this work, below, we identify some considerations that may affect the true value of the $1$ in $5$ number if accounted for.
\begin{enumerate}
  \item Closer-orbiting planets are more sensitive to $J_2$ sweeping resonances than further-orbiting planets. Because closer-orbiting planets are easier to detect via transit, systems with closer-orbiting planets will be overrepresented in the transit catalogue compared to the true, underlying distribution. Accounting for this would decrease the $1$ in $5$ rate.
  
  \item Three-planet systems seem ubiquitous in our Galaxy, with an estimated $30\%$ of sun-like stars hosting a system of $\sim3$ planets with sub-$400$ day periods \citep[][]{Zhu+18Architecture}.  As this is an expected value, higher multiplicity systems are also likely and have been detected (for example, the $7$-planet TRAPPIST-1 system \citep{Gillon+16} or the $8$-planet Kepler-90 system \citep[e.g.,][]{Cabrera+14, Shallue+18}). A larger number of planets in a system increases the possibility of a resonance, and finding previously undetected planets in observed systems is a common occurrence. Therefore,  some three-planet systems may be four-planet systems or higher. Accounting for this would increase the $1$ in $5$ rate.  
  
  \item Our sample of $31$ systems only includes systems where all three planets transit. This biases our sample towards systems that have low present-day mutual inclinations. Any system that was sufficiently misaligned by evolving $J_2$ from a transiting configuration such that it does not transit today could never be included in our sample. In other words, our selection criteria can only produce systems where the induced misalignments are low enough for the planets to remain transiting. Accounting for this would increase the $1$ in $5$ rate.
  
  \item In our calculations we adopted the upper estimate of the $J_2$ evolution depicted in Figure \ref{fig:j2evolution}. Not all stars will be born on the upper end of the initial spin distribution, and thus some systems may not undergo sweeping resonance that they might otherwise experience if the host star was born spinning faster. Accounting for this would reduce the $1$ in $5$ rate.
  
\end{enumerate}

{The third consideration naturally prompts the following questions}. {Shouldn't the sweeping resonant systems we found have misaligned themselves out of their co-transiting configurations?} Is it surprising that {as many systems as we found lie} within the sweeping resonance region {that misaligns orbits}? As an example, consider the Kepler-9 system. Kepler-9 d, the inner planet, orbits with $a/R_\star=5.54$. Assuming for simplicity that the outer two planets have transit impact parameters of $0$, then Kepler-9 d would need a mutual inclination of over $\arctan (R_\star/a) \approx 10^\circ$ with the outer planets in order to escape transit. Figure \ref{fig:alignmentchange} indicates that to become that inclined, the initial mutual inclination between the outer planets would have to be $\sim 4^\circ$ or greater. The distribution of mutual inclinations for observed {co-transiting} planets is well-modeled by a Rayleigh Distribution with $\sigma=1^\circ-2^\circ$ \citep[see e.g.,][]{Fang+12,Fabrycky+14}. {If the true distribution of initial mutual inclinations is similar to this model,} it makes sense that Kepler-9 d still transits today. Even though it has likely been misaligned from its companions via $J_2$ sweeping resonances, it would require an initial mutual inclination between outer planets of $\sim\,4^\circ$ to prevent the inner planet from transiting (see Figure \ref{fig:alignmentchange}). Of the systems we found, Kepler-9 is the most easily inclined, so for none of the systems identified is it particularly surprising that all three planets transit.

This naturally brings to mind the possibility of systems where the inner planet is so greatly misaligned from its companions that it no longer transits. These systems would likely present as two-planet transiting systems, as the outer planets align as a consequence of misaligning the inner (see the negative slopes of the blue curves in Figure \ref{fig:alignmentchange}). 
However, two effects add together to potentially limit the occurrence rate of such systems: {1)} high required mass ratios and {2)} configurational restrictions. The first effect is relatively simple: the higher the mass ratio between the inner and outer planets, the more misalignment is produced. 
For example, the mass ratio between Kepler-9's inner and middle planets is $\sim 5$, which enables the steep slope of $\Delta i_{01}^{mut}$ with respect to $i_{mut,12,init}$. While a mass-ratio of $5$ is relatively small compared to the $\sim 300$ between Earth and Jupiter, mass ratios in observed{, compact, multi-transiting} systems tend to be relatively small \citep[e.g.,][]{Weiss+18Peas}. 

The second effect limiting how often systems can be misaligned out of transit is configurational. {As mentioned, }further-orbiting inner planets require less misalignment in order to escape transit. But the minimum semimajor axis ratios for sweeping resonance to occur increase as the inner planet is further from the host star. This is because sweeping resonance requires commensurability in magnitude between planet-planet precession and $J_2$ precession, which can't happen if all planets are far from the star, but close to one another. This means that the further the inner planet orbits from the star, the larger the semimajor axis ratios between the middle and outer planets and the inner planet need to be (see for example the difference in sweeping resonance locations for Kepler-619 compared to Kepler-616 in Figure \ref{fig:9grid}). The further the inner planet orbits, the more physically isolated it must be from its companions in order to experience sweeping resonance, and such systems don't appear to be common.

The impact of sweeping resonances via $J_2$ also poses an interesting possibility. The size of the sweeping region in semimajor axis ratio space (see Figure \ref{fig:9grid}) depends on the maximum value of $J_2$ that the star experiences. This means that the present-day mutual inclinations between planets depend on the maximum $J_2$ achieved, meaning that the maximum spin of the star (and therefore the maximum $J_2$) could be constrained by present-day planetary configurations. Because stellar spins as functions of time tend to be relatively indistinguishable after $\sim 1\,$Gyr regardless of initial spin \citep[see, e.g.,][]{Barnes+03,Hartman+10,Delorme+11, Meibom+15}, this dynamical channel could provide a way to probe the initial conditions of billions of years-old stars. Simultaneously, the trends in Figure \ref{fig:alignmentchange} show a remarkable dependence {between} mean mutual inclination at late times and initial mutual inclination. {This relationship may} allow some constraints to also be put on the maximum mutual inclination between outer planets during the early years of the system's life--when $J_2$ was moving through its high values that drive the resonances. {A straightforward implementation of this is present in Figure \ref{fig:alignmentchange}, represented by the blue dotted line: blue dots above this line should not present as co-transiting systems and are thus forbidden. }

We note that if a system has parameters today that lie within the sweeping resonance region, that is not a guarantee that it actually underwent sweeping resonance in its past. 
One potential concern is whether the true stellar mass or radius deviates significantly from the values in the literature--it is noted in Figure \ref{fig:schematic} that both $M_\star$ and $R_\star$ affect where the sweeping resonance lies. {Moreover, if $J_2$ decrease begins earlier than we assume, the disk may still be present and the planets may still be migrating, meaning that the starting $J_2$ of the post-migration epoch would be lower. We discuss this possibility in detail in Appendix \ref{sec:timing}.} We explore observed systems to indicate how common sweeping-resonant architectures may be rather than attempting to claim with certainty which of these $31$ systems did or did not undergo sweeping resonance via $J_2$. Making such claims is possible; however, we do not do so here.

\section{Conclusion}\label{sec:conclusion}

In this work, we explored the implications of the evolution of stellar oblateness ($J_2$) on systems of multiple planets during the first Gyr, showing that $J_2$ evolution can bring nonresonant systems into and out of secular resonance (``sweeping resonance"). {While sweeping resonances were pointed out before in the context of $J_2$ in multi-planet systems \citep[e.g.,][]{Brefka+21,Faridani+23,Spalding+18}, here we focus on the implications for observed systems. }

{As a case study, we focus on three-planet systems. 
As time goes by, magnetic braking slows the spin of the star, reducing $J_2$. A three-planet system, initially not in resonance, can be captured in a Laplace-Lagrange resonance at a certain $J_2$. If the inner planet is as-massive or less-massive than its companions, the mutual inclination between the outer two planets is damped out as the system enters resonance. In that case, the mutual inclination between the outer pair and the inner planet increases to conserve angular momentum. {This can produce large mutual inclinations between the inner planet and the outer pair.}

{Whether or not sweeping resonances from $J_2$ occur {at all} depends on the semimajor axes and masses of the planets. In Figure \ref{fig:smaStrucure}, we show the relationship between the semimajor axis ratios in the system and the subsequent inclination induced on the inner planet by sweeping resonances from evolving $J_2$. A large portion of the parameter space exhibits sweeping resonances when $a_1/a_0$ and $a_2/a_0$ are both large{--a much larger portion of the space than is produced without evolving $J_2$.} 
Similarly, when $a_0$ decreases, the rate of $J_2$ precession increases relative to planet-planet precession, so more of the parameter space exhibits sweeping resonances when $a_0$ is small. 
{The mass ratios between the planets also affect where the resonances are located, but the sensitivity is small, with mass changes up to a factor of $2$ or $3$ not necessarily affecting whether a system enters resonance.}

To determine how common sweeping resonances from evolving $J_2$ are, we explored the population of observed, three-planet systems. 
{We found that $\sim 20\%$ of a sample of three-planet all-transiting systems have mass and semimajor axis ratios that indicate they underwent sweeping resonances early in their lives.}
In Figure \ref{fig:alignmentchange}, we show that the ability of sweeping resonances to misalign the inner planet from its companions is dependent on the initial mutual inclination between the outer two planets. Moreover, for all six sweeping-resonant systems we {identified}, {they} would require very large {initial} outer planet mutual inclinations to make the inner, transiting planet no longer transit{. As such high initial mutual inclinations are unlikely, that these systems are observed in co-transiting configurations is unsurprising.}
% --consistent with observations.  
Our findings have significant implications for the connection between exoplanet systems' initial conditions and their present configurations. {A fraction of systems experiencing sweeping resonances of near $20\%$ constitutes a generic, ubiquitous phenomenon.}

\section*{Acknowledgements}

The authors would like to thank Songhu Wang for useful discussions. This research has made use of the  \citet{exoplanetarchive}, which is operated by the California Institute of Technology, under contract with the National Aeronautics and Space Administration under the Exoplanet Exploration Program. T.H.F. and S.N thank Howard and Astrid Preston for their generous support. G.L. is grateful for the support from NASA 80NSSC20K0641 and 80NSSC20K0522. M.R. thanks the Heising-Simons Foundation for their support through Grant \#2023-4478.

\begin{appendix}

\section{The Impact of Lying Near a Mean-Motion Resonance}\label{sec:mmrs}

Of the six sweeping-resonant systems shown in Figure \ref{fig:9grid}, two of them have planets that lie near mean-motion resonances (MMRs). The two outer planets of Kepler-9 are approximately $1\%$ wide of a $2:1$ commensurability, and the planets of Kepler-487 lie near a $32:5:2$ commensurability. {The outer planet's orbit is much closer to the $16:1$ commensurability with the inner planet than the $5:2$ with the middle planet}. Mean-motion resonances occur on orbital timescales, and are therefore not accounted for in the secular model used heretofore, {requiring further investigation}. 

To verify that MMRs are not at play in these systems, we integrate Kepler-9 and Kepler-487 using different values of the outer planet semimajor axis {and randomized inner planet initial inclination}. This allows us to quantify how close these systems are to actually being in an MMR and to what extent that changes their behavior. We integrate using the N-body code REBOUND and use the add-on pack REBOUNDx with its implemented gravitational harmonics to incorporate a changing values of $J_2$ using the WHFast integrator \citep{rebound, reboundx, reboundwhfast}. The timestep for the WHFast integrator was set to be $3\%$ of the shortest orbital period in the system. {For Kepler-9 (Kepler-487), the outer planet's semimajor axis was varied between a period ratio with the middle (inner) planet of $1.97-2.03$ ($1.59-16.1$); the initial mutual inclination between outer planets was fixed to $2.5^\circ$ ($4.5^\circ$) for all integrations; the inner planet's initial inclination was randomized uniformly between $0-3^\circ$ ($0-3^\circ$); and all other orbital elements were held constant between integrations. The integrations were run for $5\times 10^6$ years with a $J_2$ evolution rate of $10$ times what is shown in Figure \ref{fig:j2evolution}. \citet{Faridani+23} found that such increased $J_2$ evolution does not qualitatively affect the results and only minorly affects them quantitatively. Even at the increased evolution rate, the timescale of $\dot{J_2}$ is still much slower than the Laplace-Lagrange timescales between the planets.} 

We show the results of these integrations in Figure \ref{fig:K9MMRs}, where we plot the period ratio between the outer two planets versus $\Delta i _{mut}$ over the integrations. Both $\Delta i _{mut,01}$ (red) and $\Delta i _{mut,12}$ (blue) are shown. {Both systems display no effect from the MMRs at their observed period ratios.}
{For Kepler-9, }wide {of resonance }(where {the planets are} observed to lie today) and short of the $2:1$ resonance, the systems display similar behavior, producing similar mutual inclination excitation between the inner planet and the outer pair. However, on-resonance, the system exhibits chaotic behavior, where the inclinations and eccentricities  of the planets vary within a broader range (consistent with observations; see e.g. \citet{Rice+2023}). {For Kepler-487, the outer and inner planets lie very close to a $16:1$ ratio. However, the effect of the commensurability is negligible, with changes to the outer planet period not having an effect on the system's evolution. This non-effect persists even when the $16:1$ ratio is near-exact. Moreover, the outer two planets of Kepler-487 are too distant from the $5:2$ commensurability for it to have an effect.}

{While neither Kepler-9 nor Kepler-487 lie close enough to their respective commensurabilities to impact the conclusions drawn from the secular calculations, this is not the case for all systems. As can be seen from the left panel of Figure \ref{fig:K9MMRs}, a $2:1$ commensurability between outer planets can have significant effect on the change on $\Delta i_{mut, 01}$ caused by evolving $J_2$. For the integrations near the commensurability, the evolution became chaotic. 
However, that an orbit is chaotic does not mean that secular resonances via evolving $J_2$ have no effect. Even in the chaotic regime, the mutual inclination between inner and middle planets almost always increases dramatically as $J_2$ passes through the secular resonant value (though the chaos occasionally drives the inner and middle two planets to temporarily align again, reducing the value of $\Delta i_{mut, 01}$). Because these MMR configurations are chaotic, this may also drive the inner planet's inclination even higher, potentially systematically removing such systems from transiting configurations, and therefore removing them from our sample.}

\begin{figure}
\centering
    \includegraphics[width=\linewidth]{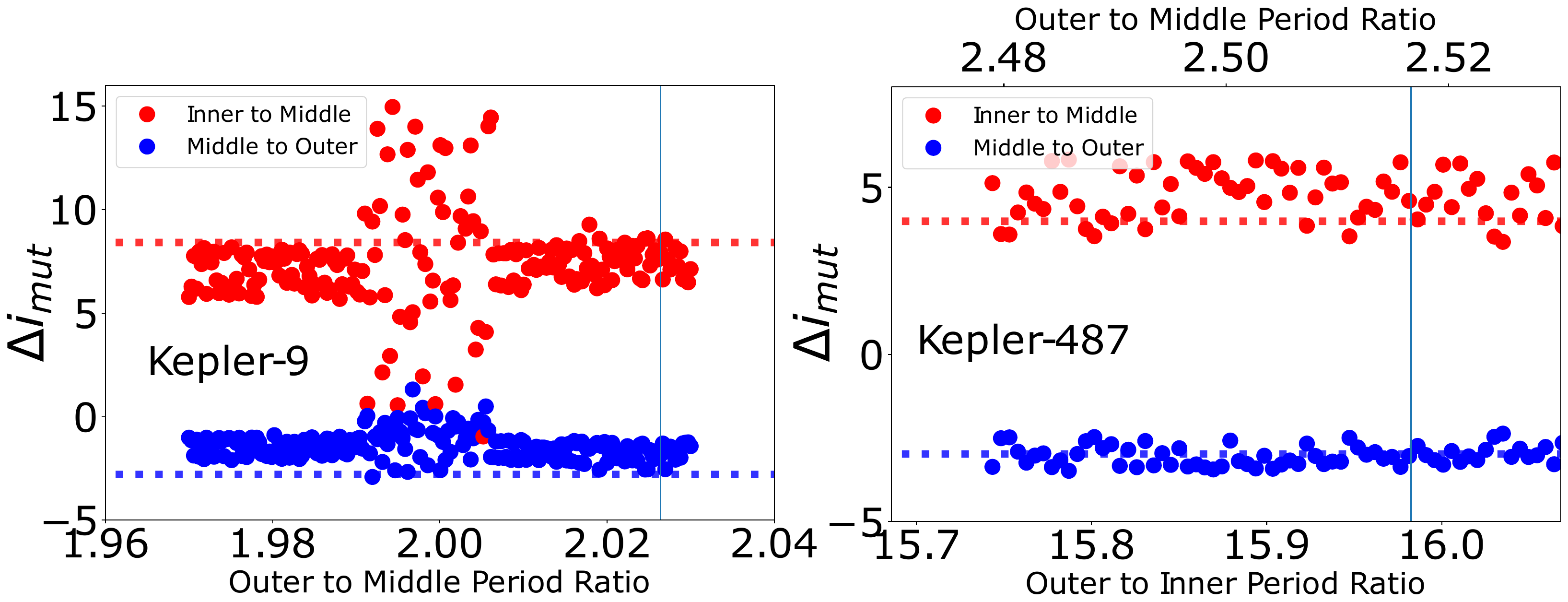}
    \caption{
    Relation between period ratio of outer two planets and $\Delta i_{mut}$ for Kepler-9 (left) and relation between period ratio of outer two planets and outer/inner for Kepler-487. The changing of the period ratio was performed by changing the outer planet's semimajor axis. The solid light blue lines represents the observed period ratios. The dotted red and blue lines represent the values of $\Delta i_{mut}$ predicted by the fits for Kepler-9 and Kepler-487 shown in Figure \ref{fig:alignmentchange}. 
    }
    \label{fig:K9MMRs}
\end{figure}

%good

\section{Sweeping Resonance from $J_2$, Planetary Migration, and the Disk}\label{sec:timing}

\begin{figure}
\centering
    \includegraphics[width=0.5\linewidth]{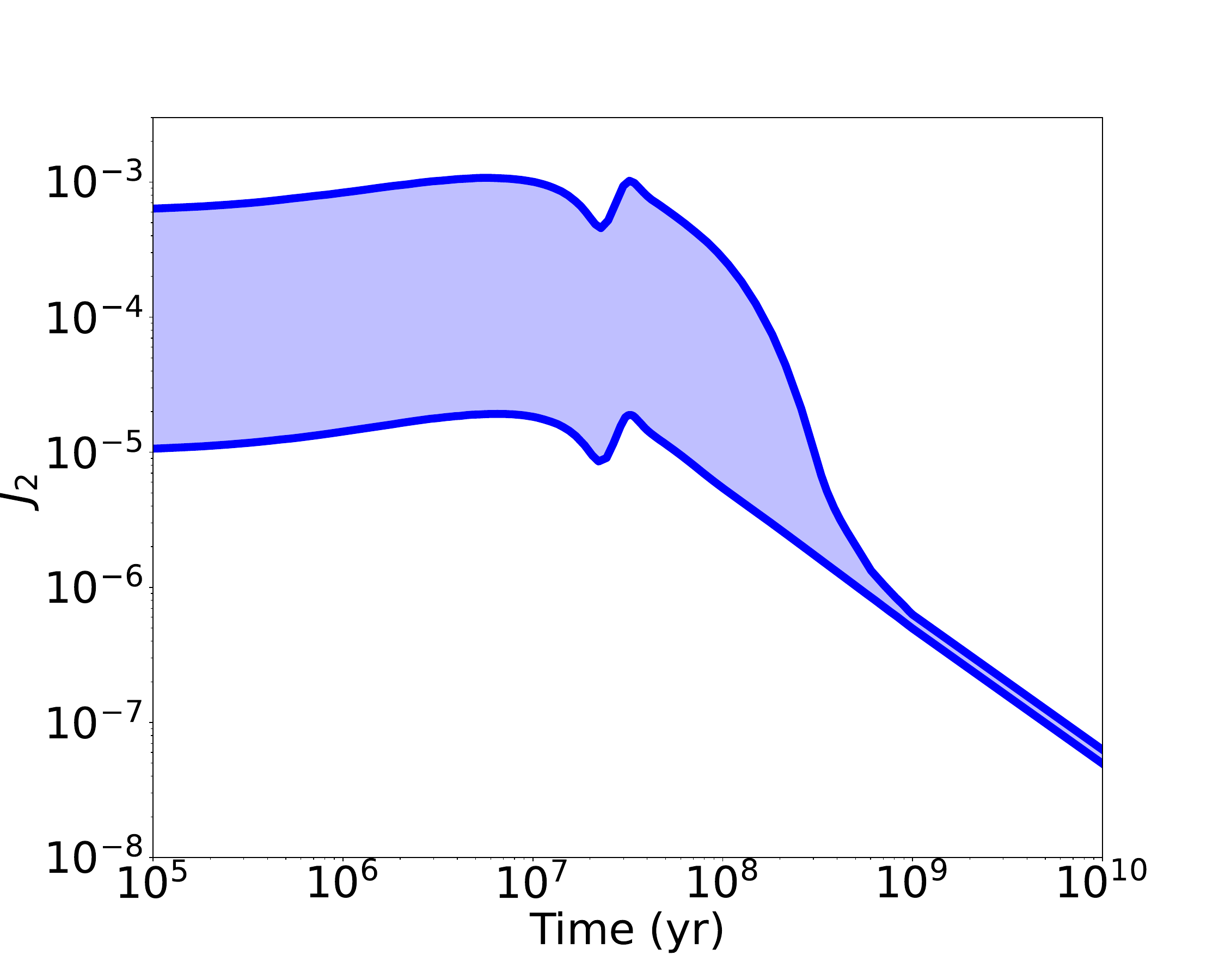}
    \caption{
    Model of $J_2$ evolution for a $1.1\, M_\odot$ star. Stellar parameters were calculated by a MESA model (the same model as Figure \ref{fig:j2evolution}), but spin evolution was calculated by the wind model given in \citet{Matt+15}.
    }
    \label{fig:J2Matt15}
\end{figure}

{The times we find these resonances occurring are relatively early in the system's life ($<50$ Myr). During this same period or earlier, the protoplanetary disk is evaporating and the planets may be migrating. The protoplanetary disk introduces its own potential, and its evaporation may prompt resonances of its own \citep[e.g.,][]{Heppenheimer+80, Ward+81,Nagasawa+03, Petrovich+20, LiLai+23, Zanazzi+24}. Planetary migration would cause a system's location in $a_1/a_0$--$a_2/a_0$ space to vary, potentially interfering with the sweeping secular resonances from evolving $J_2$, and any mean-motion resonances the system passes through could have a significant impact as well. }

{Neither of these possibilities are taken into account in our model, which assumes that the system initial conditions may be sculpted by these processes, but that they are finished making significant impact by the time $J_2$ begins evolving in our simulations at $t=10\,$Myr. While protoplanetary disks typically evaporate in less than $10\,$ Myr, older disks have been discovered, indicating the disk lifetime distribution may be wider than previously thought \citep[see e.g.][]{Silverberg+20+olddisks,Pfalzner+24+disklifetimes}. Therefore, it is necessary to consider when $J_2$ begins to evolve and whether that happens while the disk is still present.}

{The $10\,$Myr delay shown in Figure \ref{fig:j2evolution} is of somewhat arbitrary duration. However, the existence of a period where $J_2$ is roughly constant before it starts decreasing due to stellar spin-down is not arbitrary. For example, \citet{Becker+20} show in their Figure 5 the $J_2$ evolution of a $0.686\,M_\odot$ star where $J_2$ remains constant for the first $80\,$Myr of the star's life. They calculate this using \citet{Baraffe+15}'s stellar models and \citet{Matt+15}'s wind model. Using that wind model but using a MESA model for a $1.1\,M_\odot$ star produces a $\sim 60\,$ Myr lag before the $J_2$ drops below its initial value \citep[][]{Paxton2011, Paxton2013, Paxton2015, Paxton2018, Paxton2019, Jermyn2023}. This alternative $J_2$ evolution is shown in Figure \ref{fig:J2Matt15}. After $60\,$Myr, the disk is almost certainly no longer causing significant precession and migration will have stopped. Therefore, we assume disk effects and migration need not be included in our simulations. If $J_2$ begins to decrease early, then the initial $J_2$ of the post-migration epoch will be lower than assumed here.}

\section{The Sample of 31 three-planet all-transiting systems}

\begin{sidewaystable*}
% \tiny
\hspace{-2.5cm}
\centering
\begin{tabular}{lccccccccccc}
% \hspace{-1cm}
\hline \hline Name &  $M_\star$ ($M_\odot$)              & $R_\star$ ($R_\odot$)          & Star Reference & $a_0 (\mathrm{au})$ & $a_1 (\mathrm{au})$ & $a_2 (\mathrm{au})$ & $m_0 (M_\oplus)$ & $m_1 (M_\oplus)$ & $m_2 (M_\oplus)$ & Planet Reference(s)\\
% \hline Kepler-619      & $1.09\,M_{\odot}$ & $1.11\,R_{\odot}$ & \citet{Morton+16} & $\textbf{0.023}$ & $\textbf{0.062}$ & $\textbf{0.103}$ & $\textbf{1.21}$ & $\textbf{5.40}$ & $\textbf{11.68}$ & \citet{Morton+16} \\
% USP & $10^{-6}$ & $0.03$ & $10^{-3}$ & $10^{-3} \mathrm{rad}$ & $1$ \\
%system name          star mass          star radius           star cite            a0                a1                    a2               m1               m2                  m3                planet cites
% Kepler-487      & $0.91\,M_{\odot}$ & $0.88\,R_{\odot}$ & \citet{Morton+16} & $ {0.023}$ & $ {0.062}$ & $ {0.103}$ & $ {1.21}$ & $ {5.40}$ & $ {11.68}$ & \citet{Morton+16} \\
Kepler-9  & $1.07$  & $1.02$  &  \citet{Torres+11} & $ {0.027}^\dagger$ & $ {0.143}^\star$ & $ {0.230}^\star$ & ${3.342}^\dagger$ & $\textbf{43.4}^\star$ & $\textbf{29.9}^\star$ & $^\dagger$\citet{Torres+11}; $^\star$\citet{Borsato+19} \\
Kepler-619  & $1.09$  & $1.11$  &  \citetalias{Morton+16} & $ {0.022}^\dag$ & $ {0.062}^\dag$ & $ {0.103}^\star$ & $3.596^\dag$ & $14.72^\dag$ & $22.56^\star$ & $^\dag$\citetalias{Morton+16}; $^\star$\citet{Stassun2019} \\
Kepler-487  & $0.91$  & $0.88$  &  \citetalias{Morton+16} & $ {0.033}^\dag$ & $ {0.115}^\star$ & $ {0.213}^\star$ & $12.06^\dag$ & $106.3^\star$ & $11.43^\star$ & $^\dag$\citetalias{Valizadegan+22}; $^\star$\citetalias{Morton+16} \\
K2-198  & $0.8$  & $0.76$  &  \citet{Hedges+19} & $ {0.040}$ & $ {0.069}$ & $ {0.120}$ & $2.364$ & $9.882$ & $22.72$ & \citet{Hedges+19} \\
Kepler-529  & $1.07$  & $1.14$  &  \citetalias{Morton+16} & $ {0.031}^\dag$ & $ {0.109}^\dag$ & $ {0.217}^\star$ & $4.543^\dag$ & $5.422^\dag$ & $12.19^\star$ & $^\dag$\citetalias{Morton+16}; $^\star$\citetalias{Valizadegan+22} \\
EPIC 249893012  & $1.05$  & $1.70^+$  &  \citet{Hidalgo+20} & $ {0.046}$ & $ {0.124}$ & $ {0.215}$ & $\textbf{8.75}$ & $\textbf{14.67}$ & $\textbf{10.18}$ & \citet{Hidalgo+20} \\
HD 22946  & $1.1$  & $1.12$  &  \citet{Garai+23} & $ {0.051}$ & $ {0.091}$ & $ {0.264}$ & $2.124$ & $9.205$ & $10.95$ & \citet{Garai+23} \\
HD 28109  & $1.26$  & $1.45$  &  \citet{Dransfield+22} & $ {0.170}$ & $ {0.309}$ & $ {0.406}$ & $18.49$ & $7.943$ & $5.681$ & \citet{Dransfield+22} \\
K2-19  & $0.88$  & $0.82$  &  \citet{Petigura20} & $ {0.034}$ & $ {0.074}$ & $ {0.097}$ & $1.289$ & $32.4$ & $10.8$ & \citet{Petigura20} \\
K2-219  & $1.02$  & $1.19$  &  \citetalias{Mayo+18} & $ {0.048}$ & $ {0.069}$ & $ {0.098}$ & $2.071$ & $2.425$ & $10.75$ & \citetalias{Mayo+18} \\
K2-352  & $0.98$  & $0.95$  &  \citet{deLeon+21} & $ {0.046}$ & $ {0.079}$ & $ {0.117}$ & $2.155$ & $4.908$ & $8.615$ & \citet{deLeon+21} \\
K2-37  & $0.9$  & $0.85$  &  \citet{Sinukoff+16} & $ {0.051}$ & $ {0.065}$ & $ {0.110}$ & $3.194$ & $11.89$ & $11.76$ & \citet{Sinukoff+16}  \\
K2-58  & $0.89$  & $0.85$  &  \citet{Crossfield+16} & $ {0.035}$ & $ {0.069}$ & $ {0.151}$ & $3.243$ & $11.43$ & $3.700$ & \citet{Crossfield+16} \\
K2-80  & $0.9$  & $0.86$  &  \citet{Crossfield+16} & $ {0.059}^\dag$ & $ {0.134}^\dag$ & $ {0.177}^\star$ & $2.559^\dag$ & $5.489^\dag$ & $10.60^\star$ & $^\dag$ \citet{Crossfield+16}; $^\star$ \citetalias{Mayo+18} \\
% Kepler-100  & $1.08$  & $1.49$  &  \citet{Marcy+14} & $ {0.072}$ & $ {0.109}$ & $ {0.216}$ & $\textbf{7.34}$ & $\textbf{7.05}$ & $\textbf{3.0}$ & \citet{Marcy+14} \\
Kepler-1073  & $0.92$  & $1.12$  &  \citet{KepDR25} & $ {0.035}$ & $ {0.048}$ & $ {0.080}$ & $2.636$ & $3.342$ & $9.035$ & \citet{KepDR25} \\
% Kepler-1311  & $1.05$  & $1.4$  &  \citet{Morton+16} & $ {0.036}^\dag$ & $ {0.099}^\dag$ & $ {0.674}^\star$ & $1.624^\dag$ & $1.689^\dag$ & $82.15^\star$ & $^\dag$\citet{Morton+16}; $^\star$\citetalias{Valizadegan+22} \\
% Kepler-1468  & $1.04$  & $1.05$  &  \citet{Morton+16} & $ {0.046}^\dag$ & $ {0.080}^\dag$ & $ {0.144}^\star$ & $1.406^\dag$ & $3.807^\dag$ & $15.98^\star$ & $^\dag$\citet{Morton+16}; $^\star$\citetalias{Valizadegan+22} \\
Kepler-1530  & $0.92$  & $0.88$  &  \citetalias{Morton+16} & $ {0.035}^\dag$ & $ {0.058}^\dag$ & $ {0.083}^\star$ & $3.342^\dag$ & $3.391^\dag$ & $15.59^\star$ & $^\dag$\citetalias{Morton+16}; $^\star$\citetalias{Valizadegan+22} \\
Kepler-18  & $0.97$  & $1.11$  &  \citet{Cochran+11} & $ {0.044}$ & $ {0.075}$ & $ {0.117}$ & $\textbf{6.9}$ & $\textbf{17.3}$ & $\textbf{16.4}$ & \citet{Cochran+11} \\
Kepler-27  & $0.95$  & $0.85$  &  \citetalias{Morton+16} & $ {0.066}^\dag$ & $ {0.117}^\star$ & $ {0.189}^\star$ & $11.60^\dag$ & $21.16^\star$ & $28.92^\star$ & $^\dag$\citetalias{Valizadegan+22}; $^\star$\citet{Steffen+12} \\
Kepler-271  & $0.9$  & $0.87$  &  \citetalias{Morton+16} & $ {0.057}^\dag$ & $ {0.071}^\star$ & $ {0.090}^\star$ & $0.362^\dag$ & $0.882^\star$ & $1.656^\star$ & $^\dag$\citetalias{Morton+16}; $^\star$\citet{Rowe_2014} \\
Kepler-289  & $1.08$  & $1.0$  &  \citet{Schmitt+14} & $ {0.212}$ & $ {0.328}$ & $ {0.504}$ & $\textbf{7.3}$ & $\textbf{4.0}$ & $\textbf{132.0}$ & \citet{Schmitt+14} \\
Kepler-30  & $0.99$  & $0.95$  &  \citet{Sanchis-Ojeda+12} & $ {0.185}$ & $ {0.300}$ & $ {0.534}$ & $11.3$ & $640.0$ & $23.09$ & \citet{Sanchis-Ojeda+12} \\
Kepler-31  & $1.21$  & $1.22$  &  \citet{Fabrycky+12} & $ {0.158}$ & $ {0.254}$ & $ {0.411}$ & $34.55$ & $32.63$ & $20.35$ & \citet{Fabrycky+12} \\
Kepler-431  & $1.07$  & $1.09$  &  \citet{Everett+15} & $ {0.071}$ & $ {0.084}$ & $ {0.104}$ & $0.518$ & $0.373$ & $1.289$ & \citet{Everett+15} \\
Kepler-450  & $1.19$  & $1.64$  &  \citet{Stassun2019} & $ {0.079}$ & $ {0.128}$ & $ {0.193}$ & $0.647$ & $11.04$ & $40.92$ & \citet{VanEylen+15} \\
Kepler-51  & $1.04$  & $0.93$  &  \citet{Masuda+14} & $ {0.251}$ & $ {0.384}$ & $ {0.509}$ & $\textbf{2.1}$ & $\textbf{4.0}$ & $\textbf{7.6}$ & \citet{Masuda+14} \\
Kepler-549  & $0.9$  & $0.85$  &  \citet{Thompson+18} & $ {0.157}$ & $ {0.228}$ & $ {0.445}$ & $22.81$ & $10.71$ & $13.11$ & \citet{Thompson+18} \\
% Kepler-60  & $1.04$  & $1.26$  &  \citet{Jontof-Hutter+16} & $ {0.073}$ & $ {0.085}$ & $ {0.103}$ & $\textbf{4.19}$ & $\textbf{3.85}$ & $\textbf{4.16}$ & \citet{Jontof-Hutter+16} \\
% Kepler-603  & $1.01$  & $1.01$  &  \citetalias{Morton+16} & $ {0.066}$ & $ {0.149}$ & $ {0.498}$ & $1.968$ & $11.23$ & $44.67$ & \citetalias{Morton+16} \\
Kepler-616  & $0.98$  & $0.97$  &  \citetalias{Morton+16} & $ {0.090}^\dag$ & $ {0.267}^\star$ & $ {0.391}^\dag$ & $9.832^\dag$ & $11.25^\dag$ & $16.56^\star$ & $^\dag$\citetalias{Morton+16}; $^\star$\citetalias{Valizadegan+22} \\
Kepler-770  & $0.94$  & $0.92$  &  \citetalias{Morton+16} & $ {0.024}$ & $ {0.049}$ & $ {0.136}$ & $1.560$ & $2.272$ & $9.035$ & \citetalias{Morton+16} \\
KOI-7892  & $1.15$  & $1.24$  &  \citet{Stassun2019} & $ {0.099}^\dag$ & $ {0.205}^\star$ & $ {0.365}^\star$ & $13.49^\dag$ & $21.98^\star$ & $9.646^\star$ & $^\dag$\citetalias{Valizadegan+22}; $^\star$\citet{Wang+15} \\
TOI-125  & $0.86$  & $0.85$  &  \citet{Nielsen+20} & $ {0.051}$ & $ {0.081}$ & $ {0.137}$ & $\textbf{9.5}$ & $\textbf{6.629}$ & $\textbf{13.6}$ & \citet{Nielsen+20} \\
TOI-2076  & $0.82$  & $0.77$  &  \citet{Osborn+22} & $ {0.087}$ & $ {0.139}$ & $ {0.196}$ & $10.38$ & $17.21$ & $15.24$ & \citet{Osborn+22} \\

% \citet{Torres+11}
% \citet{Borsato+19}

\hline
\end{tabular}
\caption{Parameters adopted for the 31-system sample. All semimajor axes were derived from orbital periods rather than reported semimajor axes in listed references, to ensure that period ratios are maintained. Bolded masses were adopted from references, and unbolded masses were derived from the mass-radius relation derived in \citet{M-RRelationBaron+23}. $^+$: See discussion of EPIC 249893012's radius in Section \ref{sec:sweeping}. \citet{Morton+16}, \citet{Valizadegan+22}, and \citet{Mayo+18} are abbreviated as Mo16, Va22, and Ma18 respectively.}
\label{tab:allParams}
\end{sidewaystable*}

\end{appendix}

%star mass, star radius, star reference sma1, sma2, sma3, m1, m2, m3, planet reference

%%%%%%%%%%%%%%%%%%%%%%%%%%%%%%%%%%%%%%%%%%%%%%%%%%



\bibliographystyle{aasjournal}
\bibliography{kozai, paperexo} % if your bibtex file is called example.bib

%%%%%%%%%%%%%%%%%%%%%%%%%%%%%%%%%%%%%%%%%%%%%%%%%%

%%%%%%%%%%%%%%%%% APPENDICES %%%%%%%%%%%%%%%%%%%%%

%%%%%%%%%%%%%%%%%%%%%%%%%%%%%%%%%%%%%%%%%%%%%%%%%%

%%%%%%%%%%%%%%%%% APPENDICES %%%%%%%%%%%%%%%%%%%%%

%%%%%%%%%%%%%%%%%%%%%%%%%%%%%%%%%%%%%%%%%%%%%%%%%%

\bibliographystyle{aasjournal}

\end{document}